\documentclass[a4paper, fleqn]{cas-dc}

\usepackage{longtable}
\usepackage{makecell} % For wrapping text in cells
\usepackage{subcaption} % for caption in subfigures
\usepackage{tikz}
\usepackage{tabularx}
\usepackage{tcolorbox}
\usepackage{nameref}
\usepackage{multicol}

\usepackage[authoryear]{natbib}

\def\tsc#1{\csdef{#1}{\textsc{\lowercase{#1}}\xspace}}
\tsc{WGM}
\tsc{QE}
\tsc{EP}
\tsc{PMS}
\tsc{BEC}
\tsc{DE}

\newcommand{\updated}[1]{%
  {\color{black}%
    #1%
  }%
}

\begin{document}
\let\WriteBookmarks\relax
\def\floatpagepagefraction{1}
\def\textpagefraction{.001}
\shorttitle{An SLR on Runtime Composition in Dynamic SoSs}
\shortauthors{Ashfaq~et~al.}

\title [mode = title]{Runtime Composition in Dynamic System of Systems: A Systematic Review of Challenges, Solutions, Tools, and Evaluation \updated{Methods}}

\author[1]{Muhammad Ashfaq}[orcid=0000-0003-1870-7680]
\cormark[1]
\ead{muhammad.m.ashfaq@jyu.fi}
\credit{Conceptualization of this study, Methodology, Software}

\author[2]{Ahmed R. Sadik}[orcid=0000-0001-8291-2211]
\ead{ahmed.sadik@honda-ri.de}
\credit{Data curation, Writing - Original draft preparation}

\author[1]{Teerath Das}[orcid=0000-0003-2024-6545]
\ead{teerath.t.das@jyu.fi}
\credit{Data curation, Writing - Original draft preparation}

\author[3]{Muhammad Waseem}[orcid=0000-0001-7488-2577]
\ead{muhammad.waseem@tuni.fi}
\credit{Data curation, Writing - Original draft preparation}

\author[1]{Niko M\"{a}kitalo}[orcid=0000-0002-7994-3700]
\ead{niko.k.makitalo@jyu.fi}
\credit{Data curation, Writing - Original draft preparation}

\author[1]{Tommi Mikkonen}[orcid=0000-0002-8540-9918]
\ead{tommi.j.mikkonen@jyu.fi}
\credit{Supervision}

\affiliation[1]{organization={Faculty of IT, University of Jyv\"askyl\"a},
                postcode={40014}, 
                city={Jyv\"askyl\"a},
                country={Finland}}

\affiliation[2]{organization={Honda Research Institute Europe},
                postcode={63073}, 
                city={Offenbach},
                country={Germany}}
\affiliation[3]{organization={Faculty of Information Technology and Communications, Tampere University},
                postcode={33100}, 
                city={Tampere},
                country={Finland}}

\cortext[cor1]{Corresponding author}

\begin{abstract}
\noindent \textbf{Context:} 
\updated{Modern Systems of Systems (SoSs) increasingly operate in dynamic environments (e.g., smart cities, autonomous vehicles) where \textit{runtime composition}---the on-the-fly discovery, integration, and coordination of constituent systems (CSs)---is crucial for adaptability. 
% Despite growing interest, research on enabling these capabilities remains fragmented. 
% Despite growing interest, research on enabling runtime composition in SoSs remains fragmented.  
Despite growing interest, the literature lacks a cohesive synthesis of runtime composition in dynamic SoSs.
}

\noindent \textbf{Objective:} 
\updated{
% This Systematic Literature Review (SLR) synthesizes research on runtime composition in dynamic SoSs to identify core challenges, solution strategies, supporting tools, evaluation methods, and open gaps.
% This study synthesizes research on runtime composition in dynamic SoSs, identifying core challenges, solution strategies, supporting tools, and evaluation methods.
This study synthesizes research on runtime composition in dynamic SoSs and identifies core challenges, solution strategies, supporting tools, and evaluation methods.
}

\noindent \textbf{Methods:} 
% We systematically screened 1,774 studies published between 2019 and 2024, selecting 80 primary studies for thematic analysis.  
\updated{We conducted a Systematic Literature Review (SLR), screening 1,774 studies published between 2019 and 2024 and selecting 80 primary studies for thematic analysis (TA).}

\noindent \textbf{Results:} 
\updated{
% XX studies were finally selected and the key results are: 
Challenges fall into four categories: modeling and analysis, resilient operations,  system orchestration, and heterogeneity of CSs.  
Solutions span seven areas: co-simulation and digital twins, semantic ontologies, integration frameworks, adaptive architectures, middleware, formal methods, and AI-driven resilience.  
% The tooling landscape is dominated by service-oriented architectures and simulation platforms, though integration across toolchains and standardized benchmarks is limited.  
Service-oriented frameworks for composition and integration dominate tooling, while simulation platforms support evaluation.
Interoperability across tools, limited cross-toolchain workflows, and the absence of standardized benchmarks remain key gaps.
% Evaluation approaches vary across simulation, implementation, and human-centered studies, with applications in smart cities, healthcare, defense, and industrial automation.  
Evaluation approaches include simulation-based, implementation-driven, and human-centered studies, which have been applied in domains such as smart cities, healthcare, defense, and industrial automation.
}

\noindent \textbf{Conclusions:} 
\updated{
% The review highlights unresolved tensions---autonomy versus coordination, the modeling-reality gap, and socio-technical integration---and calls for standardized evaluation metrics, scalable decentralized architectures, and cross-domain frameworks. 
% The taxonomies and insights provided in this paper aim to guide researchers toward AI-driven adaptation and benchmarking, while supporting practitioners in designing resilient and interoperable SoSs.  
The synthesis reveals tensions, including autonomy versus coordination, the modeling-reality gap, and socio-technical integration.
It calls for standardized evaluation metrics, scalable decentralized architectures, and cross-domain frameworks. 
% The taxonomies and insights provided aim to guide researchers in developing solutions and practitioners in enabling effective runtime composition in dynamic SoSs.
The analysis aims to guide researchers and practitioners in developing and implementing dynamically composable SoSs.
}

\end{abstract}

\begin{highlights}
% \item Unified taxonomy of runtime composition challenges in dynamic Systems of Systems

% \item Service-oriented architectures prevail despite fragmented tool ecosystems

% \item Gaps found in AI-driven adaptation, standardized benchmarking, and socio-technical factors

\item Reviews runtime composition in the dynamic System of Systems (SoS)
\item Identifies key challenges: modeling, orchestration, resilience, heterogeneity
\item Synthesizes seven solution strategies from recent SoS literature
\item Maps tools and evaluation methods used in runtime SoS research
\item Reveals gaps in integration, benchmarking, and socio-technical alignment

\end{highlights}


\maketitle

\section{Introduction}
\label{sec:introduction}

In today's interconnected world, modern systems increasingly operate in dynamic and unpredictable environments.
In this context, the concept of a System of Systems (SoS) has emerged as an innovative paradigm for designing complex software ecosystems that comprise independently developed and managed constituent systems (CSs)\footnote{SoS refers to a single System of Systems and SoSs to multiple Systems of Systems; CS refers to a single Constituent System and CSs to multiple Constituent Systems.}. 
These CSs must interoperate and collaborate to achieve overarching goals beyond any individual system's scope.
Examples of SoSs include patient monitoring systems and smart cities, each integrating and coordinating diverse CSs to achieve overarching objectives.

% \textbf{Motivation}: As systems evolve according to Lehman's laws of software evolution~\citep{lehman1979understanding}, modern SoSs are no longer static entities. 
% They must adapt, evolve, and reconfigure in real time, often with shorter lifecycles (composing temporarily to achieve specific goals before disintegrating). 
% This shift brings forward the notion of \textit{runtime composition}, which is the ability to dynamically discover, integrate, and coordinate CSs at runtime, frequently in response to partially known or completely unforeseen environmental conditions.

\paragraph{Motivation.}
\updated{
% Modern SoSs operate in dynamic and unpredictable environments, requiring continuous adaptation and real-time reconfiguration. This has brought forward the notion of \textit{runtime composition}, which is the ability to dynamically discover, integrate, and coordinate CSs at runtime, often in response to partially known or unforeseen environmental conditions. 
Due to volatile environments, modern SoSs require continuous adaptation and real-time reconfiguration. Consequently, \textit{runtime composition}---the ability to dynamically discover, integrate, and coordinate CSs at runtime---has become a crucial mechanism for responding to partially known or unforeseen environmental conditions.
}

% While composing independent, heterogeneous, and distributed systems is manageable at the design phase, achieving this composition at runtime presents significant challenges. 
% In addition, reconfiguring a system based on pre-stored environmental inputs with known CSs is feasible, but adapting to new, unknown CSs at runtime remains problematic. 
% This challenge represents a significant difficulty in SoSE: integrating potentially heterogeneous, independent, distributed systems in a cooperative environment~\citep{9593274}.

\updated{Runtime composition with unknown CSs presents a fundamental challenge in System of Systems Engineering (SoSE)~\citep{incose, 7479441}. 
% While SoSs can reconfigure based on pre-stored environmental inputs with known CSs, discovering and integrating with unforeseen CSs that are independent, heterogeneous, and distributed (Section~\ref{sec:background}) during operation remains largely unsolved~\citep{9593274}.
While SoSs can reconfigure based on pre-stored environmental inputs with known CSs, the challenge of discovering and integrating unforeseen CSs that are independent, heterogeneous, and distributed (Section~\ref{sec:background}) during operation remains largely unsolved~\citep{9593274}.
}

% Runtime composition is particularly relevant for industrial applications at [\textit{author's anonymous institute}], where autonomous vehicle fleets require dynamic adaptation to changing environments and mission parameters. 
% Our work on connected vehicle ecosystems faces precisely the integration challenges identified in this review, where runtime composition capabilities could enhance vehicle-to-infrastructure coordination and fleet management systems.

\updated{This capability is particularly relevant for industrial applications such as urban mobility and energy SoSs.
Projects such as the Honda eVTOL initiative~\citep{evtol}, connected vehicle fleets~\citep{sose_2025, ashfaq2024holon}, large-scale trials of bidirectional electric vehicle charging~\citep{FCR}, and vehicle-to-building energy management~\citep{stadie2021v2b} illustrate the potential of runtime composition.
If achieved, it can enhance interoperability, vehicle-to-infrastructure coordination, and fleet management.
Beyond the mobility sector, open-source frameworks like CityLearn~\citep{city_learn} exemplify the broader industrial and societal relevance of dynamic SoS solutions.
Such initiatives highlight the practical challenges and emerging solutions in runtime composition, underscoring the need for a systematic review to consolidate current knowledge and guide future research.
}

\paragraph{Objectives.} 
\updated{
% Despite growing interest in runtime composition for dynamic SoSs, the literature remains fragmented and lacks a cohesive synthesis. 
% This Systematic Literature Review (SLR) addresses this gap by:
% (1) identifying key challenges to runtime composition, 
% (2) analyzing proposed solutions and frameworks,
% (3) evaluating the maturity of supporting tools and evaluation methods, and
% (4) uncovering trends and open research questions (RQs). 
Despite growing interest in runtime composition in dynamic SoSs (Section~\ref{sec:dynamic_sos}), the literature lacks a cohesive synthesis. This Systematic Literature Review (SLR) addresses this gap by: 
(1) identifying key challenges to runtime composition, 
(2) analyzing proposed solutions and frameworks, 
(3) identifying supporting tools, and 
(4) evaluation methods.
}

\paragraph{Contributions.} 
\updated{
This study makes the following key contributions through a systematic and in-depth synthesis of the literature on runtime composition in dynamic SoSs:

\begin{itemize}
    \item \textit{A unified taxonomy of runtime composition challenges}, categorized into four dimensions: modeling and analysis, resilient operations, system orchestration, and heterogeneity.
    \item \textit{A comprehensive mapping of solution strategies}, encompassing digital twins, semantic ontologies, integration frameworks, formal methods, and artificial intelligence (AI) driven resilience mechanisms.
    % \item \textit{An evaluative analysis of supporting tools}, identifying both mature platforms and critical integration limitations.
    \item \textit{An analysis of supporting tools and platforms}, highlighting both well-established environments and current limitations in tool support.
    \item \textit{A structured assessment of evaluation methods}, including commonly used methods and application domains.
    \item \textit{The identification of critical research gaps}, including the need for standardized benchmarking, unified modeling of emergent behavior, scalable decentralized control, cross-domain interoperability, and vendor-agnostic middleware.
\end{itemize}
% Collectively, these contributions address the fragmented nature of existing research, provide an integrated foundation for future work, and support both researchers and practitioners in advancing runtime composition within the SoSE landscape.
These contributions provide a foundation for future work and guide researchers and practitioners in advancing dynamic SoSs.
}

\updated{
\paragraph{Structure.} The remainder of this study is structured as follows. 
Section~\ref{sec:background} presents the background on dynamic SoSs and runtime composition. 
Section~\ref{sec:methodology} describes the SLR methodology, detailing the planning, execution, and reporting phases. 
Section~\ref{sec:results} reports the findings and synthesized taxonomies derived from the selected studies. 
Section~\ref{sec:discussion} interprets these findings, discussing their implications for research and practice. 
Section~\ref{sec:threats-to-validity} identifies limitations of the SLR. 
Section~\ref{sec:related-work}  situates our SLR within the broader literature. 
Finally, Section~\ref{sec:conclusion} concludes the paper and outlines directions for future research.}

\section{Background}
\label{sec:background}

% This section  introduces SoS, including its characteristics, types, and architectures, providing context for our SLR. 
This section introduces SoS, including its characteristics, types, and architectures, providing context for our SLR and motivating the problem statement.
Readers interested in a more comprehensive treatment are referred to~\citep{jamshidi2008systems} and \citep{nielsen2015systems}.

An SoS is an ensemble of operationally and managerially independent systems that achieve objectives beyond any individual CS's capabilities. 
An example of an SoS is a patient monitoring system that integrates CSs, such as information systems, measurement components, and diagnostic analysis tools~\citep{benabidallah2020simulating}. 
% While each CS exists as a system with its well-defined mission, their integration serves a broader purpose of maintaining patient stability, a goal beyond the capacity of any individual component. 
While each CS has its well-defined mission, its ensemble serves the broader purpose of maintaining patient stability, a goal beyond the capacity of any single CS.
Similarly, smart cities represent another example of SoS, where diverse systems must be prepared to join and contribute, even temporarily~\citep{teixeira2020constituent}.
% Other use cases of SoSs can be found in various domains, including smart grids, supply chains, intelligent transportation systems, emergency response networks, and healthcare infrastructures.
SoSs can be found in various domains, including smart grids, supply chains, intelligent transportation systems, emergency response networks, and healthcare infrastructures.

\subsection{SoS Characteristics}
\label{sec:sos_characteristics}

SoSs are distinguished from monolithic systems based on several  characteristics. These properties fall into two categories: CS-level characteristics and SoS-level characteristics.

CS-level characteristics include geographic distribution, as they can be physically dispersed, and two forms of independence: operational and managerial (O/M independence).
Operational independence means that each CS can function autonomously and would continue to operate even if removed from the SoS.
Managerial independence implies that each CS is governed separately, with its goals, resources, and development plans.
This O/M independence enables CSs to join, leave, and adapt within an SoS without disrupting their internal operations or management structures~\citep{song2022continuous}.
Closely related to independence is \textit{autonomy}~\citep{mittal2020autonomous}, where CSs maintain the ability to achieve goals while operating independently of external control, extending beyond mere automation to include \textit{behavioral autonomy} in both static contexts and dynamic environments. 
In this way, behavioral autonomy ensures adaptability at the SoS level, while \textit{structural autonomy} enables infrastructure to incorporate CSs dynamically.

At the SoS level, two characteristics particularly distinguish these systems: \textit{evolutionary development} and \textit{emergent behavior}. 
An SoS is rarely complete but instead undergoes dynamic evolution~\citep{9593274} in which new CSs are added, current CSs are replaced or removed, and the interdependency network can switch to a more efficient structure over time.
The interactions among CSs lead to SoS-level behaviors, both intended and unintended, not present in or predictable from individual CSs. 
For example, a traffic jam is an emergent misbehavior (unintended emergent behavior) as no single vehicle is designed to cause congestion, yet it arises from their collective interactions. In contrast, vehicle platooning is an intended emergent behavior, with cars coordinating to form convoys and optimize traffic flow~\citep{oquendo2019architecting}.

Additional characteristics proposed in the literature to further distinguish SoS from traditional systems include belonging, connectivity, diversity, networks, heterogeneity, and trans-domain operation.
However, several characteristics overlap (e.g., connectivity and network topology) and can be merged to streamline the definitional framework.

\subsection{SoS Types}
\label{sec:sos_types}

Based on their coordination mode,  \citep{isoiecieee21841} defines four SoS types: directed, acknowledged, collaborative, and virtual.
\textit{Directed SoS} features a central authority that creates and manages the SoS for a specific purpose, with CSs subordinated to the SoS goals. 
A military operation center controlling various units exemplifies this type. 
The \textit{acknowledged SoS} has recognized objectives and a designated manager, but CSs remain independently owned and managed; changes occur by cooperative agreement. A federated consortium of companies sharing data under central governance illustrates this type.
\textit{Collaborative SoS} lacks a single authority, but CSs voluntarily cooperate toward agreed goals, collectively deciding how to share services. 
Independent firms jointly forming a supply chain represent this model. 
A \textit{virtual SoS} has neither central management nor predefined purpose; large-scale behaviors emerge from loosely coupled CSs. An example is an open network of devices on the internet that self-organize to provide a service.

Except in Directed SoS, CS must balance autonomy and belonging~\citep{9593274}; designers and managers cannot unilaterally dictate when a CS joins or leaves the SoS, since each CS's O/M independence governs its  participation.

This classification represents a spectrum from most to least centrally directed, with the reverse order indicating the extent of CS coordination authority. The type of SoS fundamentally shapes how CSs are integrated to achieve overarching goals.

\subsection{SoS Architectures}

The architecture of an SoS specifies its CSs, the interactions among them, and the mediators enabling those interactions~\citep{silva2020verification}. 
Architecturally, an SoS can be viewed from two perspectives: (1) a set of abstract-level decisions encoded as variables with allowed alternatives  and (2) description models supported by architecture frameworks~\citep{9593274}.

Four primary architectural styles have emerged to address the diverse needs of different SoS contexts. A \textit{centralized architecture} features a single controller issuing commands to all CSs through top-down orchestration, simplifying coordination but introducing a single point of failure and limiting scalability. 
A \textit{hierarchical architecture} distributes control across multiple layers where higher-level systems coordinate lower-level subsystems, improving flexibility while maintaining central oversight. 
\textit{Heterarchical architectures} eliminate central control, with CSs cooperating through peer-to-peer interactions, offering fault tolerance at the cost of coordination complexity. 
The \textit{holonic architecture} models each CS as a \textit{holon}---a semi-autonomous unit functioning both as a self-contained system and as part of a larger hierarchy---enabling both bottom-up assembly and top-down decomposition of goals~\citep{blair2015holons}.

These architectural styles align with the SoS types (Section~\ref{sec:sos_types}). 
Directed SoSs employ centralized architectures to maintain tight control over CSs. 
Acknowledged SoSs generally implement hierarchical architectures that balance central coordination with local autonomy. 
Collaborative SoSs benefit from heterarchical architectures where coordination emerges through consensus rather than top-down control. 
Virtual SoSs often employ  heterarchical or holonic architectures, allowing system behavior to emerge organically without formal coordination structures.

\subsection{The Shift Towards Dynamic SoSs}
\label{sec:dynamic_sos}

Modern SoS environments require CSs to react jointly  to runtime changes from  other CSs and the external environment~\citep{wudka2020reconfiguration}. This dynamic context necessitates \emph{runtime composition}---the ability to discover, integrate, and reassemble CSs on‐the‐fly despite heterogeneity in interfaces, behaviors, and objectives. Such capabilities allow CSs to join, leave, be replaced, or be modified within the operational SoS~\citep{neto2017model}, enabling adaptation and evolutionary development~\citep{forte2022towards}.

Achieving robust runtime composition is challenging for several interrelated reasons. Most CSs have limited individual capabilities, necessitating integration to create required emergent behaviors~\citep{wudka2020reconfiguration}. These CSs are highly heterogeneous, originating from different providers, maintained by distinct entities~\citep{lopes2016sos}, and designed with specific missions rather than interoperability in mind~\citep{nadira2020towards}. They differ in protocols, data formats, workflows, and interfaces~\citep{garces2019software}. Furthermore, CSs often function as \textit{black boxes} where only interfaces are known, while internal structures and behaviors remain inaccessible~\citep{silva2020verification}.

Compounding these difficulties is uncertainty about which CSs will participate and the operational environment~\citep{oquendo2019dealing, oquendo2019coping, oquendo2020fuzzy, agarwal2023architecting}---a concern emphasized by both INCOSE~\citep{incose} and the European Commission's Cyber-Physical SoS (CPSoS) Research Agenda~\citep{7479441}.

Current engineering methodologies have limitations in adequately addressing these challenges. Existing approaches remain predominantly static, making evolution difficult once CSs become outdated~\citep{zabasta2020adaptive}. Most implementations rely on predefined adaptation strategies, which become infeasible when attempting to anticipate all possible scenarios~\citep{oquendo2019coping, lee2021generation, oquendo2020fuzzy}. 
These predefined adaptation strategies degrade performance when discrepancies exist between design-time knowledge and runtime events~\citep{lee2021generation}. 
Furthermore, several solutions take a top-down engineering approach, unsuitable for future SoSs that require bottom-up, autonomous integration~\citep{teixeira2020constituent}.

Achieving robust on‐the‐fly composition raises a spectrum of open questions ranging from translating between disparate data formats to guaranteeing safety when reconfiguring black‐box components.

\section{Methodology}
\label{sec:methodology}

% \color{blue}

Our methodology is guided by the well-established procedures of ~\cite{kitchenham2015evidence} and comprises three phases: (i) planning the review (Section~\ref{sec:review_planning}), (ii) executing the review (Section~\ref{sec:executing_review}), and (iii) reporting (Section~\ref{sec:reporting_phase}), as shown in Fig. 1.
To ensure transparency and replicability, we provide a replication package, including our search string, inclusion/exclusion steps, data extraction forms, quality assessments, and synthesized data~\cite{dataset}.

\begin{figure*}[t]
    \centering
    \includegraphics[width=\linewidth]{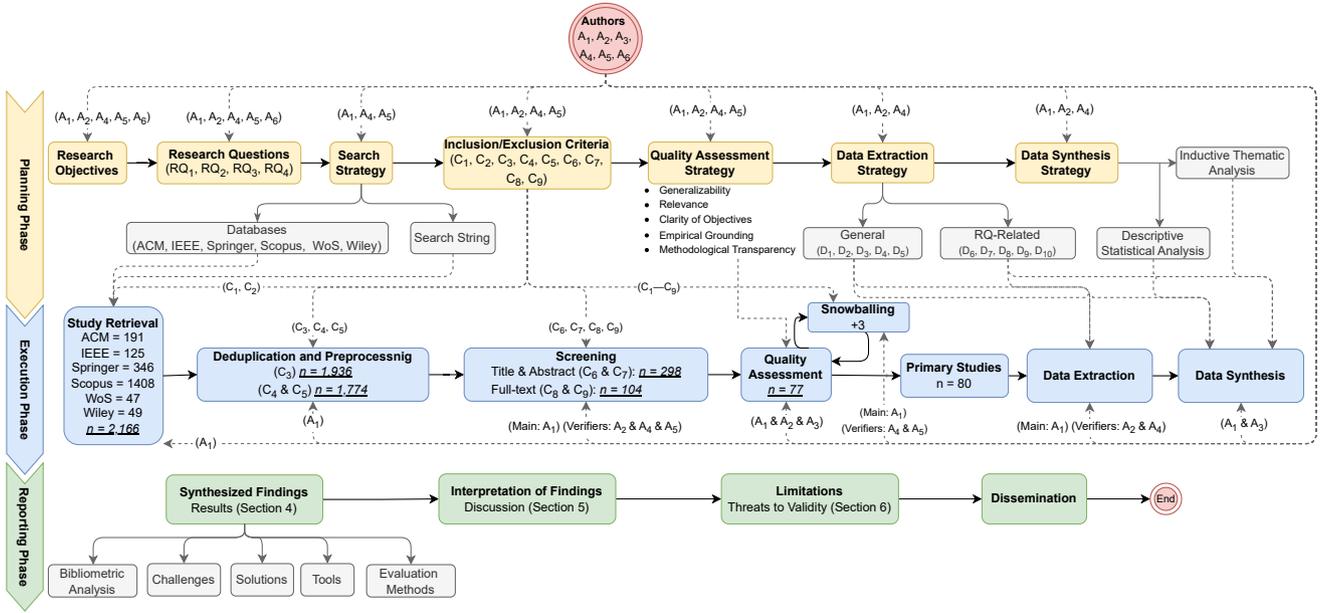}
    \caption{\updated{Methodology adopted for the SLR, illustrating its planning, execution, and reporting phases.}}
    \label{fig:methodology}
\end{figure*}

\subsection{Planning Phase}
\label{sec:review_planning}

\subsubsection{Research Objective and SLR Design}
\label{sec:research_objective}
This research aims to synthesize the literature on runtime composition in dynamic SoSs by identifying core challenges, existing solutions, supporting tools, and  evaluation methods. To achieve this, we conducted an SLR that consolidates current knowledge in the field.

While a Systematic Mapping Study (SMS) can also be used for surveying the literature~\citep{petersen2008systematic}, such studies primarily aim to provide a broad classification of research topics and trends. 
In contrast, our focused and integrative objectives require the deeper analysis and interpretive synthesis characteristic of an SLR.

\subsubsection{Research Questions}
\label{sec:research_questions}

Table~\ref{tbl:rqs} presents the four RQs and their underlying rationales, which structure and guide this SLR.

\begin{table*}[t]
\footnotesize
    \centering
    \caption{\updated{Research questions and their corresponding rationales}.}
    \label{tbl:rqs}
    % {\color{blue}
    \begin{tabular}{p{0.03\textwidth}p{0.35\textwidth}p{0.57\textwidth}}
        \hline
        \rowcolor[rgb]{0.90,0.90,0.90}
        \textbf{\#} & \textbf{Research Question} & \textbf{Rationale} \\
        \hline
        RQ\textsubscript{1} 
        & 
        % What challenges and impediments arise in enabling runtime composition among heterogeneous CSs to support self-adaptation and self-integration in SoSs?
        % What are the primary challenges in enabling runtime composition among heterogeneous CSs to realize dynamically reconfigurable and adaptable SoS?
        What are the primary challenges to achieving runtime composition in dynamic SoSs?
        &  
        % To systematically identify and classify the technical and architectural challenges to address when enabling runtime composition in SoS environments.
        % To systematically identify and classify the technical and architectural challenges that impede integrating diverse systems at runtime in SoS environments, with the goal of dynamic reconfiguration and adaptability. 
        Identifying and characterizing the key challenges offers foundational insight into what hinders the realization of runtime composition in SoS contexts. This understanding is essential for framing the research landscape and guiding the development of targeted solutions.
        \\
        RQ\textsubscript{2}  
        & 
        What solutions (approaches, conceptual frameworks, and architectural strategies) have been proposed to achieve runtime composition in dynamic SoSs?
        & 
        % To systematically analyze recent solutions addressing runtime composition among heterogeneous CSs, providing an overview of current research directions and solution spaces.
        % To systematically analyze current approaches that facilitate interoperability, dynamic integration, and reconfiguration of heterogeneous CSs, providing an overview of current research directions and solution spaces.
        A qualitative synthesis of proposed solutions can reveal underlying design philosophies, recurring assumptions, and emerging architectural patterns. This can guide future innovations and help practitioners align strategies with state-of-the-art thinking.
        \\
        % RQ3 &  
        % What are the key technical mechanisms and architectural patterns underpining successful self-adaptation and self-integration approaches in SoS?
        % What are the core technical mechanisms (e.g., semantic mediation, dynamic discovery) and strategies  underpining successful runtime composition and dynamic reconfiguration strategies in SoS?
        % & 
        % To identify and analyze core technical strategies and design patterns that enable runtime composition, revealing common solution elements and emerging technical trends across approaches
        % To uncover the core technical strategies and design patterns that address the challenges of dynamic interaction, communication, and collaboration between heterogeneous CSs, highlighting common elements and emerging trends across approaches.
        % \\\hline
        RQ\textsubscript{3} 
        & 
        % What tools and software frameworks  support self-adaptation and self-integration in SoS?
        % What tools and software frameworks are available to support the implementation of runtime composition, interoperability, and dynamic reconfiguration in SoS?
        What tools, platforms, or software frameworks have been used or proposed to support runtime composition in dynamic SoSs?
        & 
        % To identify and classify the practical tooling ecosystem that enables runtime composition in SoS, providing researchers and practitioners with an overview of available implementation resources.
        % To map the tooling ecosystem, offering researchers and practitioners insights into the available resources for designing and implementing dynamic and adaptive SoSs.
        Supporting practical tools is essential for operationalizing research ideas. This RQ provides insight into the field's technological maturity and highlights the gap between conceptual approaches and usable, integrable toolchains in real-world settings.
        \\
        RQ\textsubscript{4}  & 
        % What evaluation methods and metrics  validate self-adaptation and self-integration solutions in SoS?
        % What evaluation methodologies  assess runtime composition solutions in SoSs, particularly regarding adaptability?
        How are runtime composition approaches evaluated in the literature, and what application domains are considered?
        & 
        % To systematically analyze the evaluation landscape, including (1) assessment methods (e.g., case studies, simulations, experiments), (2) application domains and contexts used for validation, and (3) specific metrics and criteria used to measure effectiveness, thereby providing a comprehensive framework for evaluating future solutions.
        % To systematically analyze the evaluation approaches (e.g., case studies, simulations, experiments) to establish a framework for assessing the effectiveness of proposed solutions.
        This RQ explores the evaluation methods and contexts used to assess runtime composition solutions. Understanding the rigor and diversity of evaluation methods, along with the domains in which solutions are applied, helps gauge current research's maturity and practical relevance.
        \\\bottomrule
        
    \end{tabular}
    % }
\end{table*}

% \subsubsection{Search Strategy}
% \label{sec:search-strategy}

% We identified six widely used digital libraries as our search sources: ACM Digital Library, IEEE Xplore, Elsevier Scopus\footnote{Elsevier Scopus offers broader coverage compared to Elsevier ScienceDirect~\citep{ScopusVsScienceDirect}.}, SpringerLink, Wiley Online Library, and Web of Science. These databases were chosen for their comprehensive coverage of computing and engineering literature and their frequent use in similar SLRs~\citep{zhang2011empirical}. 
% After selecting the sources, we conducted an \textit{automated} search across all six libraries to retrieve relevant primary studies. No manual venue-based search or expert contact was conducted, as the selected libraries already provide broad and overlapping coverage of relevant publication venues.

% The search string was derived from the RQs and key concepts under investigation.
% We refined the search string against a manually curated reference set of 15 high-quality studies to ensure adequate recall of relevant studies. The final search string applied across all databases was:

% \texttt{(``system-of-systems'' OR ``systems-of-systems'' OR ``system of systems'' OR ``systems of systems'') \textbf{AND} \\ (``runtime composition'' OR ``dynamic reconfiguration'' OR ``self-adaptation'' OR ``self-integration'' OR \\ ``interoperability'') \textbf{AND}  (framework\textsuperscript{*} OR method\textsuperscript{*} OR tool\textsuperscript{*} OR architecture)}

\subsubsection{Search Strategy}
\label{sec:search-strategy}

Our search strategy was conducted in two complementary phases to maximize coverage of relevant primary studies.

\paragraph{Phase 1: Automated Database Search.}  
We identified six widely used digital libraries as our primary search sources: ACM Digital Library, IEEE Xplore, Elsevier Scopus\footnote{Elsevier Scopus offers broader coverage compared to Elsevier ScienceDirect~\citep{ScopusVsScienceDirect}.}, SpringerLink, Wiley Online Library, and Web of Science. These databases were chosen for their comprehensive coverage of computing and engineering literature and frequent use in similar SLRs~\citep{zhang2011empirical}.  
No manual venue-based search or expert contact was conducted, as the selected libraries already provide broad and overlapping coverage of relevant publication venues.

The search string was derived from our RQs and key concepts under investigation. 
To calibrate the string, we manually curated a baseline reference set of 13 high-quality studies, ensuring these studies represented the most relevant and influential work in the targeted research area. 
Keywords were then extracted from both the RQs and the baseline studies and combined using logical operators to form the initial search string.

% Pilot searches were conducted to test the initial string against the baseline reference set. 
% The string was iteratively refined to ensure that all 13 baseline studies were retrieved (maximizing recall) while controlling the number of irrelevant hits (maintaining precision). 

% Among the extracted terms, we included \textit{interoperability}, which was emphasized in prior work~\citep{10.1145/2465478.2465490} and in several baseline studies. 
% Interoperability challenges are fundamental to enabling runtime composition in SoS, as heterogeneous CSs must be able to interoperate to support dynamic integration and reconfiguration.

The final search string applied across all six libraries was:
\texttt{(``system-of-systems'' OR ``systems-of-systems'' OR ``system of systems'' OR ``systems of systems'') \textbf{AND} \\ (``runtime composition'' OR ``dynamic reconfiguration'' OR ``self-adaptation'' OR ``self-integration'' OR ``interoperability'') \textbf{AND} (framework\textsuperscript{*} OR method\textsuperscript{*} OR tool\textsuperscript{*} OR architecture)}

The replication package includes the derived keywords from the RQs, the baseline reference set, and the keywords extracted from the baseline reference set.

\paragraph{Phase 2: Snowballing.}  

We applied backward and forward snowballing following Wohlin's guidelines~\citep{wohlin2014guidelines} to complement the automated search. Consistent with best practice, the \textit{start set} for snowballing was defined as the set of primary studies retained after screening and quality assessment, ensuring that iterations would be based on a reliable corpus of relevant studies. All candidate studies identified through snowballing were subjected to the same inclusion, exclusion, and quality criteria as in Phase 1.

\subsubsection{Inclusion and Exclusion Criteria}
\label{sec:incl_excl_criteria}

We defined the inclusion and exclusion criteria based on insights from a pilot run of the search string to refine the search results. 
% These criteria were applied throughout the review process---from the formulation of the search string to title and abstract screening, full-text review, and snowballing (Section~\ref{sec:executing_review}). 
Table~\ref{tbl:inc_exlc_criteria} presents each criterion along with the execution phase in which it was applied.

\begin{table*}[t]
    \footnotesize
    \centering
    \caption{\updated{Inclusion and Exclusion Criteria, along with the review execution phase, where each criterion was applied.}}
    \label{tbl:inc_exlc_criteria}
    % {\color{blue}
    \begin{tabular}{p{0.01\textwidth}p{0.37\textwidth}p{0.37\textwidth}p{0.2\textwidth}}
        \hline
        \rowcolor[rgb]{0.90,0.90,0.90}
        \textbf{ID} & \textbf{Inclusion Criteria} & \textbf{Exclusion Criteria} & \textbf{Execution Phase} \\
        \hline

        C\textsubscript{1} &
        Studies must have been published within the last five years (January 1, 2019--September 22, 2024). &
        Studies published outside this timeframe. &
        Study Retrieval (Section~\ref{sec:study_retrieval})
        \\

        C\textsubscript{2} &
        Studies must be written in English. &
        Studies in other languages (e.g., Chinese, Spanish, French). &
        Study Retrieval (Section~\ref{sec:study_retrieval})
        \\

        C\textsubscript{3} &
        Studies must be unique (most recent version retained, e.g., journal extensions of conference papers). &
        Duplicate versions of the same study. &
        Deduplication and Preprocessing (Section~\ref{sec:deduplication_and_preprocessing})
        \\

        C\textsubscript{4} &
        Studies must be peer-reviewed published articles, conference papers, or book chapters. &
        Non-peer-reviewed or grey literature (e.g., blog posts), reference works, books, theses, and review articles\textsuperscript{*}. &
        Deduplication and Preprocessing (Section~\ref{sec:deduplication_and_preprocessing})
        \\

        C\textsubscript{5} &
        Studies must be published as individual research articles or book chapters. &
        Full journals, conference proceedings volumes, editorials, prefaces, and retracted articles.	 &
        Deduplication and Preprocessing (Section~\ref{sec:deduplication_and_preprocessing})
        \\

        C\textsubscript{6} &
        Studies must originate from the Computer Science or closely related engineering domains, focusing on computational, architectural, or systems aspects. &
        Studies outside the relevant domains (e.g., Business \& Management). &
        Title \& Abstract Screening (Section~\ref{sec:screening})
        \\

        C\textsubscript{7} &
        Studies must explicitly mention the SoS context in the title or abstract. &
        Studies that lack SoS context in both the title and abstract. &
        Title \& Abstract Screening (Section~\ref{sec:screening})
        \\

        C\textsubscript{8} &
        Studies must describe a solution or approach relevant to the research subject and provide sufficient methodological detail for assessment. &
        Studies that lack methodological detail or are non-substantive (e.g., vision papers, fast abstracts, opinion articles). &
        Full-text Screening (Section~\ref{sec:screening})
        \\

        C\textsubscript{9} &
        Studies must explicitly address runtime composition in dynamic SoSs or related challenges, solutions, tools, or evaluation methods. &
        Studies unrelated to runtime composition in dynamic SoS. &
        Full-text Screening (Section~\ref{sec:screening})
        \\

        \multicolumn{4}{p{0.94\textwidth}}{\footnotesize\textsuperscript{*}Review articles are not considered as primary studies; however, relevant review articles are included as related works~(Section~\ref{sec:related-work}).}
        \\\bottomrule
    \end{tabular}
    % }
\end{table*}

\subsubsection{Quality Assessment Strategy}
\label{sec:quality_assessment_strategy}

We designed a structured quality assessment rubric to ensure methodological rigor and consistency in study inclusion. 
Each study was evaluated against five predefined quality criteria:
\begin{enumerate}
    \item \textit{Relevance}:  The study addresses runtime composition or related challenges in the SoS context.
    \item \textit{Clarity of Objectives}: The research problem, aim, or hypothesis is clearly stated.
    \item \textit{Methodological Transparency}: Sufficient methodological detail is provided.
    \item \textit{Empirical Grounding}:  The study presents empirical evaluation (e.g., experiments, case studies).
    \item \textit{Generalizability }: The findings are applicable beyond a single domain or system.
\end{enumerate}

Each criterion was scored as 0, 0.5, or 1. The total score per study ranges from 0 to 5. Studies scoring $>2$ were considered to meet the quality standard and were included in the SLR.

\subsubsection{Data Extraction Strategy}
\label{sec:data_extraction_strategy}

To address the RQs, we defined a structured schema of data items (Table~\ref{tbl:data-items}). These items are grouped into two categories: (1) general study characteristics (D\textsubscript{1}--D\textsubscript{5}) capturing metadata and demographics, and (2) thematic elements (D\textsubscript{6}--D\textsubscript{10}) supporting analysis aligned with the RQs.

\begin{table*}[t]
    \footnotesize
    \centering
    \caption{Data items to be extracted from the selected studies and their corresponding RQs.}
    \label{tbl:data-items}
    \begin{tabular}{p{0.04\textwidth}p{0.18\textwidth}p{0.65\textwidth}p{0.08\textwidth}}

        \hline
        \rowcolor[rgb]{0.90,0.90,0.90}
        \textbf{Code} & \textbf{Data Item} & \textbf{Description} & \textbf{RQ} \\
        \hline
         D\textsubscript{1}  & Study ID & Unique identifier for the study to mention in the manuscript (e.g., S01, S02, ...). & Overview \\

        D\textsubscript{2}  &	Author(s) &	List of authors to find the prominent authors. & Overview \\
        
        D\textsubscript{3}  &	Publication Year  & 	For publication timeline and trend analysis.  & Overview   \\
        
        D\textsubscript{4}  &   Study Title &	Full title of the study.   & Overview  \\
        
        D\textsubscript{5}  &	Publication Type  &	Journal article, conference paper, book chapter, etc. & Overview  \\
        
        D\textsubscript{6}  &	Challenges Identified &	Specific challenges mentioned (e.g., heterogeneity, interoperability, self-integration). & RQ\textsubscript{1}  \\
        
        D\textsubscript{7}  &	Proposed Solution &	The proposed method, process, framework, or architecture. & RQ\textsubscript{2}  \\
                
        D\textsubscript{8}	& Tools \&  Technologies &	Tools, technologies, or software used to implement the solution. & RQ\textsubscript{3}     \\
        
        D\textsubscript{9} &	Evaluation Methods &	Evaluation type (e.g., case study, simulation, empirical study). & RQ\textsubscript{4}  \\
        
        D\textsubscript{10} & Application Domain & Application domains addressed in the selected studies.  & RQ\textsubscript{4}  \\

        \hline        
    \end{tabular}
\end{table*}

\subsubsection{Data Synthesis Strategy}
\label{sec:data_synthesis_strategy}

% Given the nature of the extracted data---quantitative (D1–D5) and qualitative (D6–D10)---we applied a hybrid synthesis approach.

% For D1--D5, we performed descriptive statistical analyses to capture publication trends and metadata.

% For D6--D10, we conducted a reflexive, inductive thematic analysis (TA) following Braun and Clarke's six-phase method~\citep{braun2006using, braun2019reflecting}. This method was chosen due to its suitability for synthesizing heterogeneous qualitative findings common in software engineering reviews~\citep{cruzes2011recommended}.

% We carried out open coding, iterative theme refinement, and interpretive abstraction. The final themes were structured around the RQs to ensure alignment and traceability.

We applied a hybrid data synthesis approach reflecting the nature of the extracted data. Quantitative items (D1--D5) were analyzed using descriptive statistics to identify publication trends and metadata patterns. Qualitative items (D6--D10) were synthesized using reflexive thematic analysis (TA), following Braun and Clarke's six-phase method~\citep{braun2006using, braun2019reflecting}. This approach is well-suited to synthesizing heterogeneous qualitative data common in software engineering reviews~\citep{cruzes2011recommended}.

\subsubsection{Team Roles and Responsibilities}
\label{sec:review_team_roles}

The review team's roles and responsibilities were distributed across different phases of the SLR to ensure methodological rigor and reduce potential bias. For clarity, the authors are referred to as A\textsubscript{1} through A\textsubscript{6} throughout this section.
A\textsubscript{1} served as the review lead throughout the process, consistent with Kitchenham's recommendation for PhD-led reviews~\citep{kitchenham2015evidence}.

The \textit{research objective} (Section~\ref{sec:research_objective}) and \textit{RQs} (Section~\ref{sec:research_questions}) were collaboratively formulated by A\textsubscript{1}, A\textsubscript{2}, A\textsubscript{4}, A\textsubscript{5}, and A\textsubscript{6}, drawing on their combined expertise in software and systems engineering.

The \textit{search strategy} (Section~\ref{sec:search-strategy}) was jointly designed by A\textsubscript{1}, A\textsubscript{4}, and A\textsubscript{5}. A\textsubscript{1} and A\textsubscript{4} conducted pilot searches and refined the search string. Subsequently, A\textsubscript{1} executed the final search, managed the digital library queries, and consolidated the initial dataset (Section~\ref{sec:study_retrieval}).

A\textsubscript{1}, A\textsubscript{2}, A\textsubscript{4}, and A\textsubscript{5} jointly developed the \textit{inclusion and exclusion criteria} (Section~\ref{sec:incl_excl_criteria}). \textit{Screening} (Section~\ref{sec:screening}), including \textit{title and abstract screening} and \textit{full-text screening}, was led by A\textsubscript{1}, who applied these criteria. A\textsubscript{2}, A\textsubscript{4}, and A\textsubscript{5} independently verified a subset of screening decisions. Discrepancies were resolved through discussion and consensus.

A\textsubscript{1}, A\textsubscript{2}, A\textsubscript{4}, and A\textsubscript{5} jointly developed the \textit{quality assessment strategy} (Section~\ref{sec:quality_assessment_strategy}), while A\textsubscript{1}, A\textsubscript{2}, and A\textsubscript{3} independently assessed the quality of each study in full.

The \textit{snowballing process} (Section~\ref{sec:snowballing}) was primarily conducted by A\textsubscript{1}. A\textsubscript{4} and A\textsubscript{5} independently verified the findings and supported the validation of included studies.

The \textit{data extraction} (Section~\ref{sec:data_extraction_strategy}) and \textit{synthesis strategy} (Section~\ref{sec:data_synthesis_strategy}) were co-developed by A\textsubscript{1}, A\textsubscript{2}, and A\textsubscript{4}. Following this, A\textsubscript{1} performed the initial \textit{data extraction} (Section~\ref{sec:data_extraction_and_synthesis}). A\textsubscript{2} and A\textsubscript{4} independently reviewed a subset of data items (D\textsubscript{6}--D\textsubscript{10}) that required interpretive judgment to confirm coding consistency and minimize the risk of misclassification. The data synthesis phase (Section~\ref{sec:data_extraction_and_synthesis}) was carried out through close collaboration between A\textsubscript{1} and A\textsubscript{3}, who jointly performed inductive coding and developed the thematic structure.

For verification, independent checks were performed on randomly sampled studies. Discrepancies were resolved in joint meetings, and whenever these discussions led to refinements in the criteria or assessment guidelines, the updates were consistently applied across all studies.

% Protocol Finalization and Registration  

\subsection{Execution Phase}
\label{sec:executing_review}
% This phase consists of four steps: 
% identifying the primary studies (Section~\ref{sec:primary-studies}), 
% quality assessment of the studies (Section~\ref{sec:quality-assessment}),  and
% data extraction (Section~\ref{sec:data-extraction}). 
% We explain these steps in detail as follows.

% \begin{figure}[t]
%     \centering
%     \includegraphics[width=\linewidth]{figures/Figures-studies-selection-process.pdf}
%   \caption{Overview of the review execution phase.}
%     \label{fig:executing_review}
% \end{figure}

\subsubsection{Study Retrieval}
\label{sec:study_retrieval}

The search string was adapted for each digital library to account for syntax differences and platform-specific constraints, as presented in Table~\ref{tbl:search-strings}.

\begin{table*}[!ht]
    \centering
    \caption{Data Sources and their respective search strings.}
    \label{tbl:search-strings}
    \begin{tabular}{m{0.06\textwidth}m{0.94\textwidth}}
    \hline
        \rowcolor[rgb]{0.90,0.90,0.90}
       \textbf{Database} & \textbf{Search String} \\
        \hline
        ACM & \footnotesize\texttt{[[All: ``system-of-systems''] OR [All: ``systems-of-systems''] OR [All: ``system of systems''] OR [All: ``systems of systems'']] \textbf{AND} [[All: ``runtime composition''] OR [All: ``dynamic reconfiguration''] OR [All: ``self-adaptation''] OR [All: ``self-integration''] OR [All: ``interoperability'']] \textbf{AND} [[All: framework\textsuperscript{*}] OR [All: method\textsuperscript{*}] OR [All: tool\textsuperscript{*}] OR [All: architecture]] \textbf{AND} [E-Publication Date: Past 5 years]} \\
        \hline
        IEEE & \footnotesize\texttt{(``system-of-systems'' OR ``systems-of-systems'' OR ``system of systems'' OR ``systems of systems'') \textbf{AND} (``runtime composition'' OR ``dynamic reconfiguration'' OR ``self-adaptation'' OR ``self-integration'' OR ``interoperability'') \textbf{AND} (framework\textsuperscript{*} OR method\textsuperscript{*} OR tool\textsuperscript{*} OR architecture)} \\
        \hline
        Wiley & \footnotesize\texttt{`` ``system-of-systems'' OR ``systems-of-systems'' OR ``system of systems'' OR ``systems of systems'' '' anywhere \textbf{AND} `` ``runtime composition'' OR ``dynamic reconfiguration'' OR ``self-adaptation'' OR ``self-integration'' OR ``interoperability'' '' anywhere \textbf{AND} ``framework\textsuperscript{*} OR method\textsuperscript{*} OR tool\textsuperscript{*} OR architecture'' anywhere} \\
        \hline
        Springer & \footnotesize\texttt{`(``system-of-systems'' OR ``systems-of-systems'' OR ``system of systems'' OR ``systems of systems'') \textbf{AND} (``runtime composition'' OR ``dynamic reconfiguration'' OR ``self-adaptation'' OR ``self-integration'' OR ``interoperability'') \textbf{AND} (framework\textsuperscript{*} OR method\textsuperscript{*} OR tool\textsuperscript{*} OR architecture)'; within English, (Computer Science|Engineering), 2019 - 2025 } \\
        \hline
        Scopus & \footnotesize\texttt{ALL(``system-of-systems'' OR ``systems-of-systems'' OR ``system of systems'' OR ``systems of systems'') \textbf{AND} ALL(``runtime composition'' OR ``dynamic reconfiguration'' OR ``self-adaptation'' OR ``self-integration'' OR ``interoperability'') \textbf{AND} ALL( framework\textsuperscript{*} OR method\textsuperscript{*} OR tool\textsuperscript{*} OR architecture) \textbf{AND} PUBYEAR > 2018 AND PUBYEAR < 2026 AND ( LIMIT-TO ( SUBJAREA , ``COMP'' ) OR LIMIT-TO ( SUBJAREA , ``ENGI'' ) ) AND ( LIMIT-TO ( DOCTYPE , ``cp'' ) OR LIMIT-TO ( DOCTYPE , ``ar'' ) OR LIMIT-TO ( DOCTYPE , ``ch'' ) ) AND ( LIMIT-TO ( LANGUAGE , ``English'' ) )} \\
        \hline
        WoS & \footnotesize\texttt{ALL=(``System of Systems'' OR ``systems of systems'' OR ``system-of-systems'' OR ``systems-of-systems'') \textbf{AND} ALL=(``runtime composition'' OR ``dynamic reconfiguration'' OR ``self-adaptation'' OR ``self-integration'' OR ``interoperability'') \textbf{AND} ALL=(framework\textsuperscript{*} OR method\textsuperscript{*} OR tool\textsuperscript{*} OR architecture)} \\
        \bottomrule
    \end{tabular}
\end{table*}

The search scope was set to \textit{all} available fields, targeting titles, abstracts, keywords, metadata, and other relevant fields to maximize coverage. This comprehensive approach increased the number of retrieved studies and enhanced the likelihood of capturing relevant literature across different indexing strategies.

We applied inclusion/exclusion criteria C\textsubscript{1} and C\textsubscript{2} at this phase (Table~\ref{tbl:inc_exlc_criteria}).

Executing the adapted search strings across all selected databases yielded 2,166 studies. The distribution of results per database is detailed in Figure~\ref{fig:methodology}.

\subsubsection{Deduplication and Preprocessing}
\label{sec:deduplication_and_preprocessing}

We observed duplicate studies during the search process, mainly due to WoS and Scopus indexing overlapping content from other databases.
Applying inclusion/exclusion criterion C\textsubscript{3} (Table~\ref{tbl:inc_exlc_criteria}) reduced the dataset to 1,936 studies.

Abstracts were required for the subsequent screening phase (Section~\ref{sec:screening}). For studies missing abstracts, these were manually retrieved through online searches.

In addition, inclusion/exclusion criteria C\textsubscript{4} and C\textsubscript{5} were applied at this phase (Table~\ref{tbl:inc_exlc_criteria}), further refining the dataset to 1,774 candidate studies.

\subsubsection{Screening}
\label{sec:screening}

\textit{Title \& Abstract Screening}: During this step, we reviewed titles and abstracts of the candidate studies, applying inclusion/exclusion criteria C\textsubscript{6} and C\textsubscript{7} (Table~\ref{tbl:inc_exlc_criteria}) to ensure domain relevance and SoS context. This screening reduced the dataset to 298 studies.

\textit{Full-text Screening}:
We conducted a thorough full-text review of the remaining studies, applying inclusion/exclusion criteria C\textsubscript{8} and C\textsubscript{9} (Table~\ref{tbl:inc_exlc_criteria}).
For studies without accessible full texts, we contacted the authors via \emph{ResearchGate} and excluded the studies if they were unavailable.
This phase refined the dataset to 104 studies.

\subsubsection{Quality Assessment}
\label{sec:quality_assessment_execution}

Using the predefined quality assessment rubric (Section~\ref{sec:quality_assessment_strategy}), each study was scored on a scale from 0 to 5, yielding a final set of 77 studies. 
% The quality scores are available in the replication package~\citep{dataset}.

\subsubsection{Snowballing}
\label{sec:snowballing}
% We performed both forward and backward snowballing based on references and citations.
% This process enabled us to incorporate three additional studies, leading to a final total of 107 primary studies.

Backward snowballing produced 1,832 studies, which were reduced to 1,737 unique studies after deduplication. Forward snowballing yielded 484 studies, corresponding to 440 unique studies. Together, the two directions produced 2,177 unique candidate studies.  

After applying inclusion, exclusion, and quality criteria, two studies (S78 and S80) were added to the primary studies. In addition, four backward-snowballing studies (B54, B704, B873, B924) consistently cited a key work (S79) not retrieved by the automated search. Given its repeated citation and centrality, S79 was evaluated against our criteria and included.

In this way, snowballing added three further primary studies (all via backward snowballing), while forward snowballing yielded none. Snowballing brought the total to 80 primary studies. The final set, along with cumulative quality scores and identifiers (e.g., S01, S02, …), is provided in Appendix~A. 
The complete list of studies retrieved through snowballing is provided in the replication package.

\subsubsection{Data Extraction and Synthesis}
\label{sec:data_extraction_and_synthesis}

Following the planned extraction and synthesis strategies (Section~\ref{sec:data_extraction_strategy} and \ref{sec:data_synthesis_strategy}), descriptive statistics summarized data items D\textsubscript{1}--D\textsubscript{5}. For D\textsubscript{6}--D\textsubscript{10}, reflexive TA was conducted, involving open coding, iterative theme refinement, and interpretive abstraction. We emphasized interpretive rather than objective theme identification, consistent with reflexive TA principles.

\subsection{Reporting Phase}
\label{sec:reporting_phase}

This phase encompasses the reporting, interpretation, and communication of findings. These aspects are addressed in detail in the following sections:

\begin{itemize}
    \item The reporting of results, including synthesized findings, is presented in Section~\ref{sec:results}.
    \item The interpretation of findings is discussed in Section~\ref{sec:discussion}.
    \item Study limitations are outlined in Section~\ref{sec:threats-to-validity}.
\end{itemize}

\color{black}
\section{Results}
\label{sec:results}
\updated{
We reported the results in dedicated sections aligned with our RQs (Table~\ref{tbl:rqs}). 
Section~\ref{sec:bibiometric-analysis} covers the demographic details of the selected studies.
Section~\ref{sec:challenges} and Section~\ref{sec:solutions} detail the challenges (RQ\textsubscript{1}) and proposed solutions (RQ\textsubscript{2}), respectively.
Section~\ref{sec:tooling-ecosystem} describes the tools (RQ\textsubscript{3}), and Section~\ref{sec:evaluation} outlines the evaluation methods (RQ\textsubscript{4}).
}

\subsection{Bibliometric Analysis}
\label{sec:bibiometric-analysis}

\subsubsection{Venue Distribution}
\label{sec:venue-distribution}

\begin{figure}[ht]
    \centering
    \includegraphics[width=.7\linewidth]{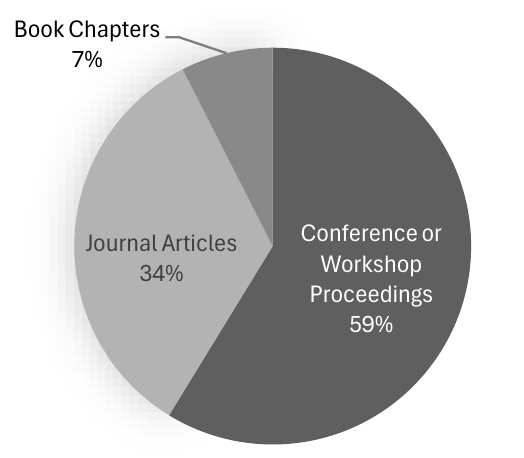}
    \caption{Distribution of publication types for the selected studies.}
    \label{fig:venue-distribution}
\end{figure}

Fig.~\ref{fig:venue-distribution} visualizes the percentage distribution of publication types of the selected studies.
Most studies appeared in conference or workshop proceedings (47), followed by journal articles (27), and a smaller number as book chapters (6).

Table~\ref{tbl:top-publication-venues} lists the top publication venues for the selected studies. The IEEE Systems Journal and SeSoS lead with four studies, followed by SESoS with three. Other notable venues include JSS, SysCon, INCOSE Journal, MILCOM, and CCIS, each contributing two studies, while the remaining venues contributed one study each. Full venue names and additional details are provided in the table.

\begin{table*}[t]
    \footnotesize
    \centering
    \caption{Most frequent individual publication venues in the selected studies.}
    \label{tbl:top-publication-venues}
    \begin{tabular}{p{0.04\textwidth}p{0.85\textwidth}p{0.05\textwidth}}
        \hline
        \rowcolor[rgb]{0.90,0.90,0.90}
        \textbf{Rank} & \textbf{Full Name} & \textbf{Studies} \\
        \hline
        \multirow{2}{*}{1} & IEEE Systems Journal & \multirow{2}{*}{4} \\
         & International Conference on System of Systems Engineering (SoSE)  & \\
        \hline
        2 & IEEE/ACM International Workshop on Software Engineering for Systems-of-Systems and Software Ecosystems (SESoS) & 3 \\
        \hline
        \multirow{9}{*}{3} & Journal of Systems and Software (JSS) & \multirow{9}{*}{2} \\
        & The Journal of International Council on Systems Engineering (Wiley INCOSE) & \\
        & IEEE International Systems Conference (SysCon) & \\
        & Communications in Computer and Information Science (CCIS) & \\
         & IEEE Military Communications Conference (MILCOM) & \\
         & IEEE/IFIP Network Operations and Management Symposium (NOMS) & \\
         & Industrial Electronics Conference (IECON) & \\
         & Reliability Engineering and System Safety & \\
        \bottomrule        
    \end{tabular}
\end{table*}

\subsubsection{Publication Trends}
\label{sec:publication-trends}

\begin{figure*}[t]
  \centering
  \begin{subfigure}[b]{0.48\textwidth}
    \includegraphics[width=\textwidth]{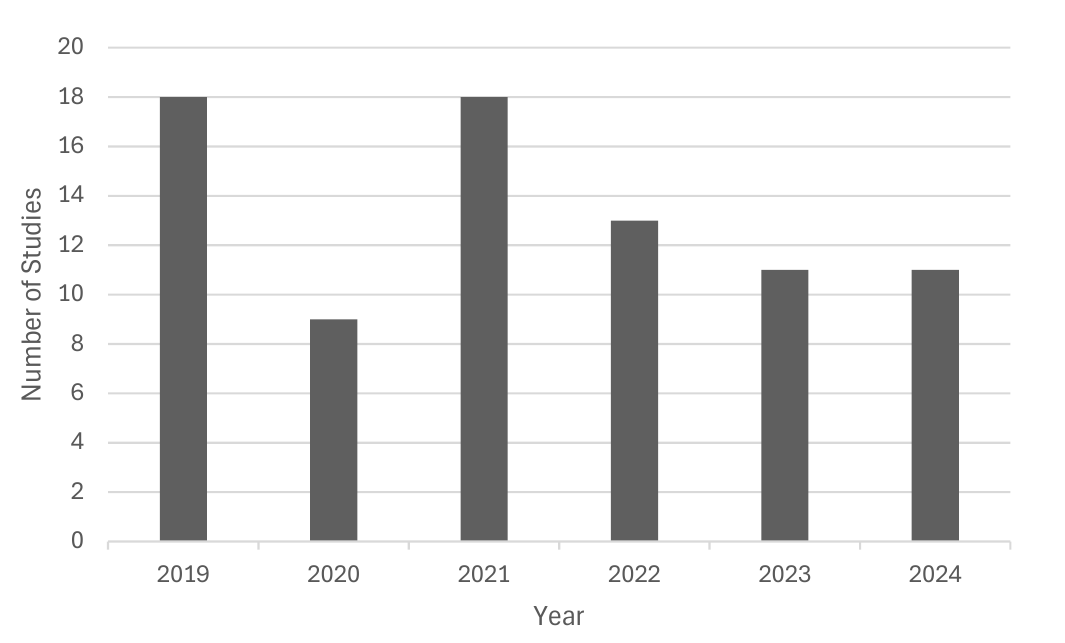} 
    \caption{Number of studies published per year.}
    \label{fig:timeline}
  \end{subfigure}
  \hfill % Add space between subfigures
  \begin{subfigure}[b]{0.48\textwidth}
    \includegraphics[width=\textwidth]{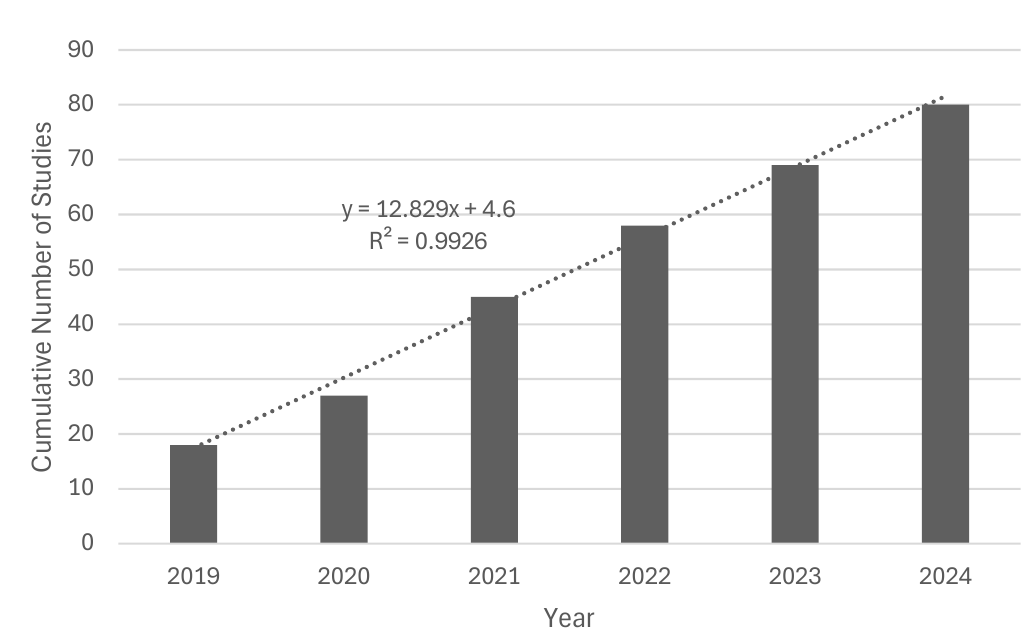} 
    \caption{Cumulative number of studies published per year.}
    \label{fig:cumulative-timeline}
  \end{subfigure}
  \caption{Publication trends of the studies selected for the SLR (2019--2024).}
  \label{fig:publication-timeline}
\end{figure*}

Fig.~\ref{fig:timeline} illustrates the publication timeline of the selected studies. Despite minor fluctuations, the overall trend remains consistent, with an average of slightly more than 10 studies published annually. Notably,  2019 and 2021 show peaks with 18 publications each, while the other years align closely with the average.

An analysis of the cumulative publications (Fig.~\ref{fig:cumulative-timeline}) reveals a strong linear growth pattern, supported by a high determination coefficient ($R^2 = 0.9926$). This growth shows that the research subject has been experiencing steady growth and increasing scholarly engagement, demonstrating the accelerating development of the research area.

\subsection{Challenges (RQ\textsubscript{1})}
\label{sec:challenges}

Table~\ref{tbl:challenges} presents our taxonomy of  challenges, organized in four themes.

\begin{table*}[t]
\footnotesize
\caption{Taxonomy of challenges faced in runtime composition of heterogeneous CSs for Dynamic  SoSs.}
\label{tbl:challenges}
\begin{tabular}{p{0.1\textwidth}p{0.12\textwidth}p{0.73\textwidth}}
\rowcolor[rgb]{0.90,0.90,0.90}
\hline
\textbf{Themes} & \textbf{Subthemes} & \textbf{Key Challenges} \\
\hline
\multirow{18}{*}{\parbox{0.2\textwidth}{Modeling \\ and Analysis}} 
& \multirow{5}{*}{\parbox{0.2\textwidth}{Unified System \\ Modeling}} 
& $\bullet$ Integrating quantitative and qualitative features in explicit resource allocation models (\hyperlink{S67}{\normalcolor S67}) \\
& & $\bullet$ Modeling heterogeneous, multi-domain systems (\hyperlink{S74}{\normalcolor S74})  \\
& & $\bullet$ Bridging micro (agent-level) and macro (organizational-level) system perspectives (\hyperlink{S62}{\normalcolor S62})   \\
& & $\bullet$ Capturing emergent behavior to reveal hidden risks and patterns (\hyperlink{S78}{\normalcolor S78})  \\
& & $\bullet$ Lack of model of computation (MoC) in control software (\hyperlink{S07}{\normalcolor S07})  \\
\cline{2-3}
& \multirow{4}{*}{\parbox{0.2\textwidth}{Multi-Domain \\ Simulation}} 
& $\bullet$ Developing standardized simulation methods for heterogeneous systems (\hyperlink{S01}{\normalcolor S01}) \\
& & $\bullet$ Creating federated simulations that synchronize across domains and incorporate physical hardware/human interactions (\hyperlink{S03}{\normalcolor S03}, \hyperlink{S47}{\normalcolor S47}) \\
& & $\bullet$ Enhancing metamodels to support dynamic system properties and discrete simulation (\hyperlink{S03}{\normalcolor S03}) \\
& & $\bullet$ Need to simulate software architecture with various requirements (\hyperlink{S57}{\normalcolor S57}) \\
\cline{2-3}
& \multirow{4}{*}{\parbox{0.2\textwidth}{Verification \& \\ Validation}} 
& $\bullet$ Minimizing verification costs in frequent reconfiguration and large model sizes (\hyperlink{S33}{\normalcolor S33}, \hyperlink{S41}{\normalcolor S41}) \\
& & $\bullet$ Ineffective verification due to modeling inaccuracy and low knowledge-sharing among CSs (\hyperlink{S41}{\normalcolor S41}) \\
& & $\bullet$ Reducing high costs and specialized knowledge requirements for physical experimental environments (\hyperlink{S20}{\normalcolor S20}) \\
& & $\bullet$ Bridging the gap between design-time and runtime artifacts (\hyperlink{S21}{\normalcolor S21}) \\
\cline{2-3}
& \multirow{4}{*}{\parbox{0.2\textwidth}{Emergence \\ Management}} 
& $\bullet$ Anticipating emergent behaviors and components due to system dynamics (\hyperlink{S79}{\normalcolor S79}) \\
& & $\bullet$ Developing methodologies to evaluate SoS resilience/performance early in lifecycle (\hyperlink{S77}{\normalcolor S77}) \\
& & $\bullet$ Creating systematic approaches for intentionally generating useful emergent properties (\hyperlink{S64}{\normalcolor S64}, \hyperlink{S79}{\normalcolor S79}) \\
& & $\bullet$ Resolving operational conflicts caused by undetected emergent behaviors (\hyperlink{S30}{\normalcolor S30}, \hyperlink{S38}{\normalcolor S38}, \hyperlink{S69}{\normalcolor S69}, \hyperlink{S76}{\normalcolor S76}) \\
\hline
\multirow{8}{*}{\parbox{0.2\textwidth}{Robust and \\ Resilient \\ Operations}} 
& \multirow{4}{*}{\parbox{0.2\textwidth}{Mission Coherence \\\& Operational \\Continuity}} 
& $\bullet$ Maintaining SoS mission coherence under external constraints (\hyperlink{S42}{\normalcolor S42}) \\
& & $\bullet$ Ensuring uninterrupted operations during system reconfigurations (\hyperlink{S37}{\normalcolor S37}) \\
& & $\bullet$ Resolving workflow indeterminacy to enable coordinated execution (\hyperlink{S50}{\normalcolor S50}) \\
& & $\bullet$ Adapting behaviors only under exceptional conditions without affecting normal operations (\hyperlink{S26}{\normalcolor S26}) \\
\cline{2-3}
& \multirow{4}{*}{\parbox{0.2\textwidth}{Resilience \&\\ Robustness}} 
& $\bullet$ Enabling fault tolerance for dynamic behavior, unpredictability, and network/component failures (\hyperlink{S15}{\normalcolor S15}, \hyperlink{S35}{\normalcolor S35}) \\
& & $\bullet$ Maintaining operational/communication independence to prevent cascading failures (\hyperlink{S11}{\normalcolor S11}) \\
& & $\bullet$ Sustaining robustness amid environmental changes, faults, and adversarial conditions (\hyperlink{S30}{\normalcolor S30}, \hyperlink{S45}{\normalcolor S45}) \\
& & $\bullet$ Implementing runtime monitoring, predictive maintenance, and adaptive resilience mechanisms (\hyperlink{S59}{\normalcolor S59}, \hyperlink{S60}{\normalcolor S60}) \\
\bottomrule

\end{tabular}
\end{table*}

\begin{table*}[t]
\footnotesize
\ContinuedFloat
\caption{Taxonomy of challenges(continued).}
\label{tbl:challenges_cont}
\begin{tabular}{p{0.1\textwidth}p{0.12\textwidth}p{0.73\textwidth}}
\rowcolor[rgb]{0.90,0.90,0.90}
\hline
\textbf{Theme} & \textbf{Subtheme} & \textbf{Key Points} \\
\hline

\multirow{21}{*}{\parbox{0.2\textwidth}{System \\ Orchestration}} 
& \multirow{4}{*}{\parbox{0.2\textwidth}{Dynamic \& \\ Scalable \\ Architectures}} 
& $\bullet$ Evolving SoS architectures beyond static models to support dynamicity and scalability (\hyperlink{S69}{\normalcolor S69}, \hyperlink{S80}{\normalcolor S80}) \\
& & $\bullet$ Managing continuously changing structures while balancing resilience with affordability (\hyperlink{S37}{\normalcolor S37}, \hyperlink{S44}{\normalcolor S44}, \hyperlink{S69}{\normalcolor S69}, \hyperlink{S75}{\normalcolor S75}, \hyperlink{S77}{\normalcolor S77}) \\
& & $\bullet$ Supporting open, vendor-independent architectures (\hyperlink{S63}{\normalcolor S63}) \\
& & $\bullet$ Balancing standardization (regulatory compliance) with flexibility (adaptability) (\hyperlink{S66}{\normalcolor S66}) \\
\cline{2-3}
& \multirow{8}{*}{\parbox{0.2\textwidth}{Dynamic \\Interoperability \\\& Coordination}} 
& $\bullet$ Ensuring an up-to-date SoS view while enabling interoperability for unstructured, knowledge-intensive processes (\hyperlink{S17}{\normalcolor S17}, \hyperlink{S76}{\normalcolor S76}). \\
& & $\bullet$ Creating on-the-fly mediators for runtime composition and mediation (\hyperlink{S54}{\normalcolor S54}) \\
& & $\bullet$ Establishing unified communications between CSs (\hyperlink{S06}{\normalcolor S06}, \hyperlink{S07}{\normalcolor S07}, \hyperlink{S19}{\normalcolor S19}, \hyperlink{S21}{\normalcolor S21}, \hyperlink{S24}{\normalcolor S24}, \hyperlink{S54}{\normalcolor S54}) \\
& & $\bullet$ Automating SoS integration while reducing costs (\hyperlink{S43}{\normalcolor S43}, \hyperlink{S52}{\normalcolor S52}, \hyperlink{S54}{\normalcolor S54}) \\
& & $\bullet$ Standardizing integration methods for heterogeneous systems (\hyperlink{S16}{\normalcolor S16}, \hyperlink{S36}{\normalcolor S36}, \hyperlink{S75}{\normalcolor S75}) \\
& & $\bullet$ Domain-agnostic coordination (\hyperlink{S66}{\normalcolor S66}) by aligning communication interfaces (\hyperlink{S12}{\normalcolor S12}, \hyperlink{S55}{\normalcolor S55}) \\
& & $\bullet$ Balancing independence with coordination requirements (\hyperlink{S44}{\normalcolor S44}, \hyperlink{S51}{\normalcolor S51}, \hyperlink{S52}{\normalcolor S52}, \hyperlink{S62}{\normalcolor S62}) \\
& & $\bullet$ Integrating autonomous AI-enabled systems within SoS frameworks (\hyperlink{S36}{\normalcolor S36}, \hyperlink{S45}{\normalcolor S45}, \hyperlink{S78}{\normalcolor S78}) \\
\cline{2-3}
& \multirow{5}{*}{\parbox{0.2\textwidth}{Dynamic \\Discovery \&\\ Composition}} 
& $\bullet$ Addressing runtime formation and discovery of potential CSs (\hyperlink{S22}{\normalcolor S22}) and their functionalities (\hyperlink{S43}{\normalcolor S43}) \\
& & $\bullet$ Identifying service-providing systems while considering quality of service (QoS) attributes (\hyperlink{S35}{\normalcolor S35}) \\
& & $\bullet$ Enabling programmatic, dynamic, and autonomous SoS construction (\hyperlink{S46}{\normalcolor S46}, \hyperlink{S52}{\normalcolor S52}, \hyperlink{S56}{\normalcolor S56}) \\
& & $\bullet$ Supporting post-deployment composition for dynamic integration (\hyperlink{S54}{\normalcolor S54}) \\
& & $\bullet$ Providing developers with assistance and automation from design-time analysis to runtime optimization (\hyperlink{S59}{\normalcolor S59}) \\
\cline{2-3}
& \multirow{3}{*}{\parbox{0.2\textwidth}{ Interaction \&\\  Resource \\ Management}} 
& $\bullet$ Enabling autonomic inter-system interaction without prior planning (\hyperlink{S27}{\normalcolor S27}) \\
& & $\bullet$ Implementing distributed, incremental resource scheduling across autonomous systems (\hyperlink{S35}{\normalcolor S35}) \\
& & $\bullet$ Managing resources and temporal constraints in evolving environments (\hyperlink{S50}{\normalcolor S50}, \hyperlink{S67}{\normalcolor S67}) \\
\cline{2-3}
& \multirow{3}{*}{\parbox{0.2\textwidth}{Control Strategies}} 
& $\bullet$ Balancing centralized control complexity with adherence to global objectives (\hyperlink{S50}{\normalcolor S50}) \\
& & $\bullet$ Enabling real-time control and task delegation in distributed environments (\hyperlink{S64}{\normalcolor S64}) \\
& & $\bullet$ Facilitating varied control strategies across independent systems (\hyperlink{S30}{\normalcolor S30}, \hyperlink{S38}{\normalcolor S38}, \hyperlink{S46}{\normalcolor S46}) \\

\hline

\multirow{13}{*}{\parbox{0.1\textwidth}{Heterogeneity of CSs}} 
& \multirow{3}{*}{\parbox{0.2\textwidth}{Component  \\Diversity}} 
& $\bullet$ Physical devices (IoT, Industry 4.0, sensors, hardware) (\hyperlink{S05}{\normalcolor S05}, \hyperlink{S07}{\normalcolor S07}, \hyperlink{S43}{\normalcolor S43}, \hyperlink{S56}{\normalcolor S56}, \hyperlink{S57}{\normalcolor S57}, \hyperlink{S60}{\normalcolor S60}, \hyperlink{S63}{\normalcolor S63}, \hyperlink{S70}{\normalcolor S70}) \\
& & $\bullet$ Digital components (abstract CSs, digital twins, multi-robot systems) (\hyperlink{S06}{\normalcolor S06}, \hyperlink{S13}{\normalcolor S13}, \hyperlink{S65}{\normalcolor S65}, \hyperlink{S68}{\normalcolor S68}) \\
& & $\bullet$ Architectural spans (legacy vs. modern frameworks) (\hyperlink{S02}{\normalcolor S02}, \hyperlink{S23}{\normalcolor S23}, \hyperlink{S27}{\normalcolor S27}, \hyperlink{S58}{\normalcolor S58}, \hyperlink{S76}{\normalcolor S76}) \\
\cline{2-3}
& \multirow{3}{*}{\parbox{0.2\textwidth}{Representation \\ Fragmentation}} 
& $\bullet$ Incompatible data formats, sources, and models (\hyperlink{S04}{\normalcolor S04}, \hyperlink{S05}{\normalcolor S05}, \hyperlink{S14}{\normalcolor S14}, \hyperlink{S44}{\normalcolor S44}, \hyperlink{S47}{\normalcolor S47}, \hyperlink{S52}{\normalcolor S52}, \hyperlink{S54}{\normalcolor S54}) \\
& & $\bullet$ Semantic mismatches in representations (\hyperlink{S34}{\normalcolor S34}, \hyperlink{S48}{\normalcolor S48}) \\
& & $\bullet$ Validation gaps between abstract and concrete implementations (\hyperlink{S06}{\normalcolor S06}, \hyperlink{S13}{\normalcolor S13}, \hyperlink{S21}{\normalcolor S21}, \hyperlink{S59}{\normalcolor S59}, \hyperlink{S60}{\normalcolor S60}) \\
\cline{2-3}
& \multirow{3}{*}{\parbox{0.2\textwidth}{Proprietary \\Constraints}} 
& $\bullet$ Vendor lock-ins and proprietary solutions (\hyperlink{S23}{\normalcolor S23}, \hyperlink{S34}{\normalcolor S34}, \hyperlink{S54}{\normalcolor S54}, \hyperlink{S63}{\normalcolor S63}, \hyperlink{S75}{\normalcolor S75}) \\
& & $\bullet$ Black-box components and defiant systems (\hyperlink{S26}{\normalcolor S26}, \hyperlink{S34}{\normalcolor S34}) \\
& & $\bullet$ Outdated protocols and legacy architectures (\hyperlink{S23}{\normalcolor S23}, \hyperlink{S58}{\normalcolor S58}) \\
\cline{2-3}
& \multirow{4}{*}{\parbox{0.2\textwidth}{Operational \\Coupling}} 
& $\bullet$ Cascading compatibility issues (\hyperlink{S16}{\normalcolor S16}, \hyperlink{S29}{\normalcolor S29}, \hyperlink{S49}{\normalcolor S49}) \\
& & $\bullet$ Mediation complexity in bridging systems (\hyperlink{S05}{\normalcolor S05}, \hyperlink{S43}{\normalcolor S43}, \hyperlink{S65}{\normalcolor S65}) \\
& & $\bullet$ Security vulnerabilities in heterogeneous networks (\hyperlink{S39}{\normalcolor S39}) \\
& & $\bullet$ Operational fragility from undocumented interactions (\hyperlink{S23}{\normalcolor S23}, \hyperlink{S58}{\normalcolor S58}) \\
\bottomrule

\end{tabular}
\end{table*}

\subsubsection{Modeling and Analysis}
\label{sec:Modeling and Analysis}
This theme reveals the limitations in  representing and analyzing complex SoS behaviors.
The \textit{Unified System Modeling} subtheme exposes critical gaps in holistic representation, particularly in integrating quantitative resource allocation with qualitative features (S67) and bridging micro-macro system perspectives (S62). Current approaches struggle with multi-domain heterogeneity (S74) and emergent behavior prediction (S78), while the absence of formal computation models in control software (S07) compounds these issues.

The \textit{Multi-Domain Simulation} subtheme highlights interoperability barriers, with studies demonstrating the need for standardized methods (S01) and federated simulations incorporating hardware/human interactions (S03, S47). Researchers emphasize requirements for dynamic property support in metamodels (S03) and architecture simulation across requirement variants (S57).

In \textit{Verification \& Validation}, key challenges include exponential cost growth during frequent reconfigurations (S33, S41) and accuracy limitations from poor knowledge sharing between CSs (S41). Physical validation environments remain prohibitively expensive (S20), while design-runtime artifact mismatches persist (S21).

Finally, the \textit{Emergence Management} subtheme emphasizes the difficulty in anticipating emergent behaviors (S79) and resolving operational conflicts caused by such behaviors (S30, S38, S69, S76). Some studies (S64, S79) point to the need for systematic approaches that can intentionally generate desired emergent properties.

\subsubsection{Robust and Resilient Operations}
This theme identifies critical vulnerabilities in maintaining SoS functionality in dynamic operational contexts.
\textit{Mission Coherence \& Operational Continuity} challenges center on preserving strategic objectives during architectural transitions (S42, S37), with particular difficulties in workflow indeterminacy resolution (S50) and exception-driven adaptation (S26).

The \textit{Resilience \& Robustness} subtheme highlights four key gaps: insufficient fault-tolerance mechanisms for network or component failures (S15, S35); cascading failure risks due to operational dependencies (S11); limited environmental adaptation capabilities (S30, S45); and challenges in integrating adaptive resilience measures such as runtime monitoring and predictive maintenance (S59, S60).

\subsubsection{System Orchestration}
\label{sec:system_orchestration}
This theme encapsulates the dynamic management and integration of  CSs.

Under the \textit{Dynamic \& Scalable Architectures} subtheme, the evolution from static to dynamic architectures that support scalability and flexibility is emphasized (S69, S80).

The \textit{Dynamic Interoperability \& Coordination} subtheme identifies challenges such as real-time mediator creation, standardizing communication protocols, and automating integration processes across domains (S06, S54).

The subtheme \textit{Dynamic Discovery \& Composition} addresses challenges in runtime formation and discovery of potential CSs (S22) and their functionalities (S43), as well as enabling programmatic, dynamic, and autonomous SoS construction (S46, S52, S56). 
S35 highlights the challenge of identifying service-providing systems while considering QoS attributes.

The \textit{Interaction \& Resource Management} subtheme emphasizes the need for mechanisms that facilitate autonomic inter-system interaction without prior planning (S27) and implement distributed, incremental resource scheduling across autonomous CSs (S35). 
Some studies highlight the challenge of managing resources and temporal constraints in evolving environments (S50, S67).

Finally, the \textit{Control Strategies} subtheme highlights the need to balance centralized control complexity with adherence to global objectives (S50), enable real-time control and task delegation in distributed environments (S64), and facilitate varied control strategies across independent CSs (S30, S38, S46). This balance is particularly challenging in integrating autonomous AI-enabled CSs within SoS frameworks (S36, S45, S78), where the goal is to align autonomous decision-making with global SoS objectives while preserving system independence.

\subsubsection{Heterogeneity of Constituent Systems}
\label{sec:Heterogeneity of Constituent Systems}
This theme addresses the multidimensional integration challenges arising from technical, organizational, and historical diversity in CSs of SoSs components. 
% The literature reveals four interconnected dimensions of heterogeneity that impede runtime composition:
The literature identifies four interconnected dimensions of heterogeneity that impede runtime composition, as outlined below.

\textit{Component Diversity} manifests through incompatible IoT ecosystems (S70), Industry 4.0 devices (S63), and specialized hardware (S05, S07) that resist standardized integration. Studies highlight particular difficulties in bridging abstract CS specifications (S06, S13) with physical implementations, especially when digital twins (S68) or multi-robot systems (S65) introduce additional interface layers. This diversity compounds architectural tensions between modern frameworks (S58) and legacy systems (S23, S58).

\textit{Representation Fragmentation} emerges from incompatible simulation models (S04, S47) and data formats (S52, S54), with studies emphasizing the cascading effects of custom information models (S34). The subtheme identifies critical gaps in toolchain interoperability (S59) and validation mismatches between design-time models and runtime behaviors, particularly in human-AI environments (S60).

\textit{Proprietary Constraints} create systemic rigidity through vendor lock-ins (S63, S75) and black-box components (S34) that obscure operational logic. Studies highlight how bespoke protocols (S23, S34) and defiant components (S26) actively resist integration efforts, while legacy architectures (S23) impose outdated communication patterns that conflict with runtime composition requirements.

\textit{Operational Coupling} reveals three compounding effects: 1) Networked dependencies (S29, S49) that propagate failures across operating systems (OS) or software-intensive SoS (SiSoS) architectures (S02, S27), 2) Secure communication impeded by protocol heterogeneity (S39), and 3) Exponential mediation costs when scaling deployments (S05, S43). Studies particularly emphasize the fragility of systems combining modern and legacy components (S16, S58).

\subsection{Solutions (RQ\textsubscript{2})}
\label{sec:solutions}

Table~\ref{tbl:solutions} presents the taxonomy of solutions proposed to enable runtime composition in dynamic SoSs. These solutions address various aspects of interoperability, integration, and reconfiguration of  CSs.

\begin{table*}[t]
\footnotesize
\caption{Taxonomy of solutions.}
\label{tbl:solutions}
\begin{tabular}{p{0.1\textwidth}p{0.12\textwidth}p{0.73\textwidth}}
\rowcolor[rgb]{0.90,0.90,0.90}
\hline
\textbf{Theme} & \textbf{Subtheme} & \textbf{Key Points} \\
\hline
\multirow{21}{*}{\parbox{0.2\textwidth}{Simulation, \\Modeling, \\and Design}} 
& \multirow{7}{*}{\parbox{0.2\textwidth}{Co-Simulation \&\\ Integrated \\ Simulation}} 
& $\bullet$ Adaptations from Smart Grid to Industry 4.0 for CS simulation and emergent behavior detection (\hyperlink{S01}{\normalcolor S01}) \\
& & $\bullet$ Cloud-based co-simulation using HLA standard and Kubernetes (\hyperlink{S48}{\normalcolor S48}) \\
& & $\bullet$ Interface between SAGM Toolbox and Mosaik (\hyperlink{S18}{\normalcolor S18}) \\
& & $\bullet$ SEAT framework: Modular, cloud-enabled, API-driven simulation testbed (\hyperlink{S08}{\normalcolor S08}, \hyperlink{S36}{\normalcolor S36}) \\
& & $\bullet$ DEDSM for integrating C model with C2 model (\hyperlink{S03}{\normalcolor S03}) \\
& & $\bullet$ MBSE and DEVS integration via SoSADL for simulation/model transformation (\hyperlink{S57}{\normalcolor S57}) \\
& & $\bullet$ Behavioral model of interoperability implemented in the agent-based simulation program MANA (\hyperlink{S49}{\normalcolor S49}) \\
& & $\bullet$ LEGO-robot-based physical exemplar for platooning and sensor/actuator data collection (\hyperlink{S20}{\normalcolor S20}) \\
\cline{2-3}
& \multirow{5}{*}{\parbox{0.2\textwidth}{Modeling}} 
& $\bullet$ Hypergraph-based modeling of SoS internal structures (\hyperlink{S11}{\normalcolor S11}) \\
& & $\bullet$ Scenario-based modeling for normal and exceptional conditions using AOP wrappers (\hyperlink{S26}{\normalcolor S26}) \\
& & $\bullet$ Escher: exposing only external behavior via a refinement-aware message bus (\hyperlink{S09}{\normalcolor S09}) \\
& & $\bullet$ Abstraction of internal details to focus on communication properties and behaviours (\hyperlink{S55}{\normalcolor S55}) \\
& & $\bullet$ A generic metamodel in Maude to control resources and behavioral features of SoS (\hyperlink{S67}{\normalcolor S67}) \\
\cline{2-3}
& \multirow{2}{*}{\parbox{0.2\textwidth}{Design-time \\Approaches}} 
& $\bullet$ ADORE: integrates Cognitive Work Analysis and Goal-Oriented Requirements Engineering to model domain analysis, collaborative tasks, and scenarios requiring adaptation strategies (\hyperlink{S38}{\normalcolor S38}) \\
& & $\bullet$ Design-time adaptation strategies seeded from a knowledge base and optimized via stochastic search (\hyperlink{S73}{\normalcolor S73}) \\
\cline{2-3}
& \multirow{6}{*}{\parbox{0.2\textwidth}{Digital Twins}} 
& $\bullet$ Digital Twins for coupling the physical and virtual worlds in SoS architecture descriptions (\hyperlink{S22}{\normalcolor S22}) \\
& & $\bullet$ Real-time simulation and monitoring with Digital Twins (\hyperlink{S61}{\normalcolor S61}) \\
& & $\bullet$ Embedding Digital Twins in synthetic environments for scenario simulation (\hyperlink{S78}{\normalcolor S78}) \\
& & $\bullet$ Leveraging digital twins for real-time simulation and monitoring (\hyperlink{S60}{\normalcolor S60}) \\
& & $\bullet$ Use of IoT technology and Digital Twins to enable dynamic reconfiguration (\hyperlink{S10}{\normalcolor S10}) \\
& & $\bullet$ Modeling CSs as DTs for runtime adaptation (\hyperlink{S40}{\normalcolor S40}) \\

\cline{2-3}
& \multirow{2}{*}{\parbox{0.12\textwidth}{Lifecycle Management}} 
& $\bullet$ SoSLM: Process-oriented lifecycle management integrating PLM and IIoT methodologies (\hyperlink{S24}{\normalcolor S24}) \\
& & $\bullet$ SoSE Lifecycle Concept: Multi-stage, bidirectional approach for smart product-based systems (\hyperlink{S10}{\normalcolor S10}) \\

\hline

\multirow{9}{*}{\parbox{0.2\textwidth}{Semantic \\Approaches}} 
& \multirow{5}{*}{\parbox{0.2\textwidth}{Mission and \\ Domain-Focused \\Ontology Design}} 
& $\bullet$ OSysRec ontology for synthesizing structural, dynamic, and management aspects of system reconfiguration (\hyperlink{S74}{\normalcolor S74}) \\
& & $\bullet$ RDF-based common models addressing model mismatches (\hyperlink{S34}{\normalcolor S34}) \\
& & $\bullet$ Ontology describing domain semantics and logical analysis (\hyperlink{S05}{\normalcolor S05}, \hyperlink{S31}{\normalcolor S31}) \\
& & $\bullet$ Linked data and ontologies for data representation and exchange (\hyperlink{S17}{\normalcolor S17}) \\
& & $\bullet$ Knowledge Intensive Process Ontology and Notation for interoperability decisions (\hyperlink{S76}{\normalcolor S76}) \\
\cline{2-3}
& \multirow{4}{*}{\parbox{0.2\textwidth}{Ontology-Driven \\ System \\ Composition \\ and Discoverability}} 
& $\bullet$ Utilizing ontologies to represent systems, enabling them to discover, understand, and compose at runtime (\hyperlink{S27}{\normalcolor S27}, \hyperlink{S46}{\normalcolor S46}) \\
& & $\bullet$ Utilizing the holon ontology to describe device properties and functionalities (\hyperlink{S43}{\normalcolor S43}, \hyperlink{S32}{\normalcolor S32}, \hyperlink{S46}{\normalcolor S46}) \\
& & $\bullet$ NLP-generated holon descriptions from publicly-available device data (\hyperlink{S32}{\normalcolor S32}) \\
& & $\bullet$ Holonic architecture enabling systems to discover, understand, and compose at runtime (\hyperlink{S46}{\normalcolor S46}) \\
\hline
\multirow{11}{*}{\parbox{0.2\textwidth}{Integration \\Approaches}} 
& \multirow{5}{*}{\parbox{0.2\textwidth}{Framework-Based  \\Integration}} 
&  $\bullet$ Conceptual Integration Framework (integrating model, tool, and execution platforms) (\hyperlink{S47}{\normalcolor S47}) \\
& & $\bullet$ Federation of rule engines for rule control and integration policies (\hyperlink{S34}{\normalcolor S34})  \\
& & $\bullet$ ISoS framework to enhance conceptual interoperability (\hyperlink{S04}{\normalcolor S04}) \\
& & $\bullet$ Use of ontologies and holons for enabling interoperability (\hyperlink{S46}{\normalcolor S46}) \\
& & $\bullet$ Using blockchain to enable trusted interoperability (\hyperlink{S64}{\normalcolor S64}) \\
\cline{2-3}
& \multirow{2}{*}{\parbox{0.2\textwidth}{Protocol-Driven \\ Integration}} 
& $\bullet$ 5G-enabled low-latency and peer-to-peer integration for heterogeneous CSs (\hyperlink{S39}{\normalcolor S39}) \\
& & $\bullet$ JSON-based protocols for CS interactions (\hyperlink{S45}{\normalcolor S45}) \\
\cline{2-3}
& \multirow{3}{*}{\parbox{0.2\textwidth}{Domain-specific \\ Integration}} 
& $\bullet$ Integration of model-based simulations with domain-specific support (\hyperlink{S44}{\normalcolor S44}) \\
& & $\bullet$ Using DSLs to cover domain-wise integration space (\hyperlink{S56}{\normalcolor S56}) \\
& & $\bullet$ ROGER framework abstracts away low-level details of CSs using domain-specific language IFP (\hyperlink{S13}{\normalcolor S13}) \\
& & $\bullet$ Seamless Integration Framework for domain-wide standardization (\hyperlink{S75}{\normalcolor S75}) \\

\bottomrule
\end{tabular}
\end{table*}

\begin{table*}[t]
\footnotesize
\ContinuedFloat
\caption{Taxonomy of solutions (continued).}
\label{tbl:solutions_cont}
\begin{tabular}{p{0.1\textwidth}p{0.12\textwidth}p{0.73\textwidth}}
\rowcolor[rgb]{0.90,0.90,0.90}
\hline
\textbf{Theme} & \textbf{Subtheme} & \textbf{Key Points} \\
\hline

\multirow{16}{*}{\parbox{0.2\textwidth}{Architectural \\ Approaches}} 
& \multirow{7}{*}{\parbox{0.2\textwidth}{Adaptive \&\\ Dynamic \\ Architectures}} 
& $\bullet$ dynADL: a lightweight ADL for dynamic architectures supporting runtime reconfiguration and component management (\hyperlink{S69}{\normalcolor S69}) \\
& & $\bullet$ Enhanced SosADL with fuzzy constructs based on Fuzzy Theory (\hyperlink{S22}{\normalcolor S22}) \\
& & $\bullet$ A metamodel and enterprise architecture for dynamic architecting of SoS configurations (\hyperlink{S44}{\normalcolor S44}) \\
& & $\bullet$ Emergence-aware architecture description method for improved design explainability (\hyperlink{S79}{\normalcolor S79}) \\
& & $\bullet$ Game-theory-based multi-domain design (\hyperlink{S29}{\normalcolor S29}) \\
& & $\bullet$ Publish-subscribe architectures for scalability, integration, and interoperability (\hyperlink{S51}{\normalcolor S51}, \hyperlink{S23}{\normalcolor S23}) \\
& & $\bullet$ Transitioning between architectures in SoSs to dynamically adapt to mission requirements through iterative reconfiguration (\hyperlink{S80}{\normalcolor S80}, \hyperlink{S27}{\normalcolor S27}) \\
\cline{2-3}
& \multirow{4}{*}{\parbox{0.2\textwidth}{Service-Oriented \\ Architectures}} 
& $\bullet$ Microservice-driven similarity analysis and GraphQL-based API unification (\hyperlink{S06}{\normalcolor S06}) \\
& & $\bullet$ MSA for modularity, scalability, and resilience (\hyperlink{S60}{\normalcolor S60}) \\
& & $\bullet$ Autonomous aggregator architecture within an SOA for smart buildings with multi-criticality policies and self-X properties (\hyperlink{S70}{\normalcolor S70}) \\
& & $\bullet$ Eclipse Arrowhead Framework (\hyperlink{S05}{\normalcolor S05}, \hyperlink{S16}{\normalcolor S16}, \hyperlink{S52}{\normalcolor S52}, \hyperlink{S12}{\normalcolor S12}, \hyperlink{S21}{\normalcolor S21}, \hyperlink{S58}{\normalcolor S58}, \hyperlink{S19}{\normalcolor S19}, \hyperlink{S63}{\normalcolor S63}) \\
\cline{2-3}
& \multirow{5}{*}{\parbox{0.2\textwidth}{Middleware \& \\Communication}} 
& $\bullet$ Architecture using TSN, MPLS, two-step admission control, distributed service discovery, and cloud resources for on-demand service instantiation (\hyperlink{S35}{\normalcolor S35}) \\
& & $\bullet$ Three-tier architecture using Discovery and Access Broker (DAB) with RM-ODP (\hyperlink{S14}{\normalcolor S14}) \\
& & $\bullet$ MediArch: three-layer mediator for interaction and control (\hyperlink{S30}{\normalcolor S30}) \\
& & $\bullet$ Hetero-Genius: middleware for runtime discovery and composition (\hyperlink{S54}{\normalcolor S54}) \\
& & $\bullet$ Employing Discrete Controller Synthesis (DCS) to automatically generate controllers for coordination (\hyperlink{S66}{\normalcolor S66}) \\

\hline

\multirow{5}{*}{\parbox{0.1\textwidth}{Formal \& \\Analytical \\Methods}}
& \multirow{2}{*}{\parbox{0.2\textwidth}{ Verification \& \\ Analytical \\ Strategies}} &
$\bullet$ Verification strategies for structural/data correctness and dynamic interactions (\hyperlink{S55}{\normalcolor S55})  \\
& & $\bullet$ Formal analytical techniques (graph theory, multi-attribute utility theory) for mission path evaluation (\hyperlink{S31}{\normalcolor S31})\\
\cline{2-3}
& \multirow{3}{*}{Hybrid Approaches} & 
$\bullet$ TRC-MM with RT-Maude for time-resource aware missions (\hyperlink{S50}{\normalcolor S50}) \\
& & $\bullet$ Continuous Verification via MAPE-K Patterns with Model Slicing (\hyperlink{S33}{\normalcolor S33}, \hyperlink{S41}{\normalcolor S41}) \\
& & $\bullet$ Maude Strategy Language for dynamic reconfigurations (\hyperlink{S42}{\normalcolor S42}) \\
\hline

\multirow{9}{*}{\parbox{0.2\textwidth}{Resilience \& \\ Fault \\Tolerance}} 
& \multirow{2}{*}{\parbox{0.2\textwidth}{Architectural \\ Reconfiguration}} 
& $\bullet$ ReViTA framework for fault tolerance via architectural reconfigurations (\hyperlink{S15}{\normalcolor S15}) \\
& & $\bullet$ CSSAD method for reconstructing SoS to restore lost operational capabilities (\hyperlink{S28}{\normalcolor S28}) \\
\cline{2-3}
& \multirow{2}{*}{\parbox{0.2\textwidth}{Hybrid\\ Reconfiguration}} 
& $\bullet$ Bottom-up fault detection with top-down reconfiguration for enhanced resilience (\hyperlink{S11}{\normalcolor S11}) \\
& & $\bullet$ Operation-loop-based mission reliability evaluation model for external shocks and topology reconfiguration (\hyperlink{S53}{\normalcolor S53}) \\
\cline{2-3}
&  \multirow{3}{*}{\parbox{0.2\textwidth}{Nature-Inspired \\ Approaches} }
& $\bullet$ Bio-inspired approach using ecological fitness principles and the DoSO metric for resilience and affordability (\hyperlink{S77}{\normalcolor S77}) \\
& & $\bullet$ Mosaic control system enabling agents for resilient adaptation (\hyperlink{S29}{\normalcolor S29}) \\
& & $\bullet$ Multi-swarm-based cooperative reconfiguration model to maximize the resilience (\hyperlink{S71}{\normalcolor S71}) \\
\cline{2-3}
& \multirow{2}{*}{\parbox{0.2\textwidth}{AI-driven \\ Approaches}} 
& $\bullet$ DRLRESF that uses GCNs and PPO for autonomous learning of resilience strategies (\hyperlink{S72}{\normalcolor S72}) \\
& & $\bullet$ Dynamically generating adaptation strategies in uncertain environment (\hyperlink{S73}{\normalcolor S73}) \\
\hline

\multirow{6}{*}{\parbox{0.1\textwidth}{Cross-Cutting \\ Approaches}} 
& \multicolumn{2}{p{0.9\textwidth}}{\raggedright
$\bullet$ Distributed Cognitive Toolkit (language-agnostic extension of CST for distributed cognitive systems) (\hyperlink{S45}{\normalcolor S45}) \\ 
$\bullet$ Integration of domain-specific architecture models with multi-agent systems for decentralized and adaptive CPS (\hyperlink{S62}{\normalcolor S62}) \\ 
$\bullet$ CorteX framework: an interoperable, modular, and extensible decentralized control system with a self-describing data model, designed for pub/sub, service-oriented applications, and demonstrated in real-world fusion hardware (\hyperlink{S02}{\normalcolor S02}, \hyperlink{S65}{\normalcolor S65}) \\ 
$\bullet$ Service-oriented approach for enabling interoperability between SOSJ and non-SOSJ through OPC UA  (\hyperlink{S07}{\normalcolor S07}) \\ 
$\bullet$ Portable data connectivity leveraging OPC-UA to encapsulate digital twins, ensuring semantic interoperability and decentralized and modular interactions for dynamic reconfiguration (\hyperlink{S68}{\normalcolor S68}) \\ 
$\bullet$ The CERBERO Project: Integrates modeling, deployment, and verification tools to enable a multi-level self-adaptation infrastructure with a hierarchical adaptation loop for CPS, supporting functional, non-functional, and repair-oriented adaptivity (\hyperlink{S59}{\normalcolor S59})
} \\ 
\bottomrule
\end{tabular}
\end{table*}

\subsubsection{Simulation, Modeling, and Design}
This theme encompasses approaches for representing, simulating, and designing SoSs, focusing on modularity and interoperability.

Solutions in subtheme \textit{Co-Simulation \& Integrated Simulation} create interfaces between different simulation platforms and testbeds (e.g., SAGM Toolbox with Mosaik; S18), and employ frameworks such as SEAT (S08, S36) to support modular, application programming interface (API) driven simulation.

Some studies demonstrate diverse \textit{Modeling} techniques, ranging from hypergraph-based representations (S11) to scenario-based frameworks (S26), with generic metamodels (S55) abstracting internal details to prioritize communication properties and emergent behaviors.

The \textit{Lifecycle Management} approaches address end-to-end SoSE through process-oriented frameworks that span design, operation, and evolution. For example, SoS lifecycle management (SoSLM) (S24) integrates Product Lifecycle Management (PLM) with industrial internet of things (IIoT) for coordinated governance in distributed environments. Meanwhile, the SoSE Lifecycle Concept (S10) introduces a multi-stage, agile methodology for smart product-based systems.

\textit{Digital Twins} couple physical and virtual systems for real-time monitoring (S61) and dynamic reconfiguration (S40). 
\textit{Design-time Approaches} like ADORE (S38) embed adaptation policies into runtime workflows, while S73 leverages knowledge bases with stochastic search optimization to generate adaptation strategies.

\subsubsection{Semantic Approaches}
This theme focuses on enhancing system understanding and interoperability through semantic technologies.

\textit{Mission and Domain-Focused Ontology Design} approaches, such as OSysRec (S74), address model mismatches by synthesizing structural, dynamic, and management aspects into resource description framework (RDF) based linked data models.

\textit{Ontology-Driven System Composition and Discoverability} techniques leverage ontologies for dynamic discovery and composition, employing holonic architectures and natural language processing (NLP) to generate CS descriptions (S32, S46).

\subsubsection{Integration Approaches}
This theme aims to provide seamless, scalable, and domain-agnostic connectivity among  CSs.

\textit{Framework-Based Integration} solutions such as infrastructure SoS (ISoS) (S04) and Holons (S46) offer high-level architectures for integrating CSs.

\textit{Protocol-Driven Integration} subtheme focuses on technical enablers for runtime composition, employing lightweight data formats like JavaScript object notation (JSON) (S45) and emerging technologies like 5G (S39) to address challenges of heterogeneity and latency.

\textit{Domain-Specific Integration} methods use domain-specific languages (DSLs) or metamodels to facilitate tighter coupling and reduced ambiguity for CSs operating within similar contexts.

\subsubsection{Architectural Approaches}

This theme comprises solutions dedicated to constructing adaptable and scalable SoS infrastructures.

Studies on \textit{Adaptive \& Dynamic Architectures} emphasize formal methods, such as lightweight ADLs, including dynaADL (S69) and fuzzy-theory-enhanced SosADL (S22), to address runtime reconfiguration and uncertainty.
Metamodels for enterprise architecture (S44) and emergence-aware methods (S79) contribute to design explainability, while game-theory applications (S29) introduce mathematical rigor into multi-domain challenges.
Practical patterns like publish-subscribe architectures (S51, S23) address scalability and interoperability, whereas architectural transitions enable SoS to dynamically adapt to evolving mission requirements through iterative reconfiguration (S80, S27).
These approaches reflect a cross-disciplinary integration that balances theoretical innovation with practical applicability.

The \textit{Service-Oriented Architectures} (SOA) subtheme highlights the role of SOA in enabling modular, scalable, and resilient runtime-composable SoSs.
Research spans specialized service implementations, such as microservice architectures (MSA) with GraphQL-based API unification (S06, S60) and domain-specific implementations (S70). It also includes standardized frameworks, exemplified by the widespread adoption of the Eclipse Arrowhead Framework (EAF) (S05, S16, S52, S12, S21, S58, S19, S63).

The subtheme of \textit{Middleware \& Communication Architectures} emphasizes real-time coordination, scalability, and dynamic resource management. It covers multi-layered approaches, such as three-tier architectures with Discovery and Access Brokers (DAB) (S14) and frameworks like MediArch (S30).
Hybrid solutions integrate Time-Sensitive Networking (TSN), Multiprotocol Label Switching (MPLS), and cloud resources (e.g., S35), while middleware frameworks like Hetero-Genius (S54) and Discrete Controller Synthesis (DCS) (S66) provide automated discovery, composition, and adaptive coordination.

\subsubsection{Formal and Analytical Methods}
This theme presents frameworks to ensure mission-critical systems' adaptability, correctness, and efficiency.

\textit{Verification Strategies and Foundational Techniques} involve methods like verification strategies (S55) and foundational theories (e.g., graph theory, multi-attribute theory (S31) to ensure analytical rigor in mission path assessment.

\textit{Hybrid Approaches} focus on integrating Time-Resource Aware Control metamodels with tools such as RT-Maude (S50) and leveraging continuous verification methods---e.g., using Monitor-Analyze-Plan-Execute over Knowledge (MAPE-K) patterns and property-based model slicing (S33, S41)---to balance static verification with dynamic control. Rewriting logic via the Maude Strategy Language (S42) also supports flexible specification of dynamic reconfigurations.

\subsubsection{Resilience and Fault Tolerance}

This theme consists of approaches that maintain SoS functionality amid failures and disruptions during dynamic composition with other CSs.

A key trend is the development of \textit{Architectural Reconfiguration Frameworks} like Reconfigurations Via Transient Architectural Configurations (ReViTA) (S15) for fault-tolerant design and CSoS autonomous decision-making (CSSAD) (S28) for post-disruption capability restoration.
These frameworks work alongside \textit{hybrid reconfiguration} models, such as bottom-up fault detection paired with top-down recovery (S11) and operation-loop-based models for mission reliability under external shocks (S53).

Some studies also highlight \textit{Nature-Inspired Principles}---ecological fitness metrics (DoSO in S77) and decentralized strategies (mosaic agents in S29, multi-swarm reconfiguration in S71)---to balance with cost-effectiveness. 

\textit{AI-Driven Techniques} enable autonomous and context-aware resilience by combining graph networks with reinforcement learning (RL) to infer resilience strategies (S72), and dynamically generating adaptation policies for uncertain environments (S73). 
These approaches illustrate a paradigm where proactive design and reactive adaptation converge to enhance system robustness in complex operational contexts.

\subsubsection{Cross-Cutting Approaches}
A subset of the reviewed solutions spans multiple themes, combining architectural, semantic, and integration strategies.
For example, the CorteX framework (S02, S65) offers an extensible, decentralized control system with a self-describing data model, built on publish-subscribe and service-oriented principles. Similarly, the CERBERO Project (S59) integrates modeling, deployment, and verification into a multi-level self-adaptation infrastructure for cyber-physical systems (CPS).
OPC UA exemplifies protocol-driven integration (S07) while extending into digital twins (S68), illustrating its cross-cutting role in semantic and architectural interoperability.
Finally, advanced cognitive and multi-agent strategies, as seen in the Distributed Cognitive Toolkit (S45) and integrating domain-specific models with multi-agent systems (S62), illustrate emerging directions in the research subject.

\subsection{Tooling Ecosystem (RQ\textsubscript{3})}
\label{sec:tooling-ecosystem}

Our synthesis reveals a diverse set of tools and technologies supporting runtime composition, interoperability, and dynamic reconfiguration in SoSs.
We organized these tools into themes and subthemes, reflecting distinct but complementary approaches to SoS design and operation (Table~\ref{tbl:tools_tech}).

\begin{table*}[t]
\footnotesize
\caption{Tooling ecosystem for runtime composition in SoSs.}
\label{tbl:tools_tech}
\begin{tabular}{p{0.16\textwidth}p{0.22\textwidth}p{0.58\textwidth}}
\rowcolor[rgb]{0.90,0.90,0.90}
\hline
\textbf{Theme} & \textbf{Subtheme} & \textbf{Tools With Study IDs} \\
\hline
\multirow{4}{*}{\parbox{0.16\textwidth}{Simulation \&\\ Scenario Testing}} 
& Co-Simulation Platforms & 
Mosaik (\hyperlink{S01}{\normalcolor S01}, \hyperlink{S18}{\normalcolor S18}, \hyperlink{S47}{\normalcolor S47}), Portico RTI (\hyperlink{S48}{\normalcolor S48} \\
\cline{2-3}
& Dynamic \& Fuzzy Model Simulation  & 
MS4ME Platform (\hyperlink{S57}{\normalcolor S57}), 
DEVS-Suite (\hyperlink{S22}{\normalcolor S22}), DEVSim++ (\hyperlink{S03}{\normalcolor S03}) \\
\cline{2-3}
& High-Fidelity Simulations (drones, urban mobility, etc)	 & 
Gazebo (\hyperlink{S40}{\normalcolor S40}, \hyperlink{S51}{\normalcolor S51}), Unity (\hyperlink{S40}{\normalcolor S40}, \hyperlink{S48}{\normalcolor S48}), NVIDIA Omniverse (\hyperlink{S68}{\normalcolor S68}), SUMO (\hyperlink{S48}{\normalcolor S48}), Unreal Engine (\hyperlink{S36}{\normalcolor S36}), Dash (\hyperlink{S40}{\normalcolor S40}), Dragonfly Simulator (\hyperlink{S26}{\normalcolor S26}), VENTOS Simulator for platooning (\hyperlink{S33}{\normalcolor S33}) \\
\cline{2-3}
& Network Simulation & 
NS3 (\hyperlink{S02}{\normalcolor S02}), OMNeT++/Mesosaurus (\hyperlink{S46}{\normalcolor S46}) \\
\cline{2-3}
& Agent-based Simulation & 
MANA (\hyperlink{S49}{\normalcolor S49}), JADE (\hyperlink{S80}{\normalcolor S80}) \\
\cline{2-3}
& Distributed Simulation & 
DIS/RPR FOM (\hyperlink{S04}{\normalcolor S04}) 
\\
\hline
\multirow{4}{*}{\parbox{0.16\textwidth}{Formal Modeling \&\\ Verification}} 
& Model-Driven Design \& Development & 
SysML (\hyperlink{S21}{\normalcolor S21}, \hyperlink{S37}{\normalcolor S37}), SGAM (\hyperlink{S18}{\normalcolor S18}, \hyperlink{S61}{\normalcolor S61}, \hyperlink{S75}{\normalcolor S75}),  PREESM (\hyperlink{S59}{\normalcolor S59}), MagicDraw (\hyperlink{S21}{\normalcolor S21}), Enterprise Architect (\hyperlink{S18}{\normalcolor S18}) \\
\cline{2-3}
& Formal Verification & 
Papyrus (\hyperlink{S55}{\normalcolor S55}), PRISM (\hyperlink{S33}{\normalcolor S33}, \hyperlink{S41}{\normalcolor S41}), 
Uppaal (\hyperlink{S55}{\normalcolor S55}), 
Maude/MaudeSL (\hyperlink{S42}{\normalcolor S42}, \hyperlink{S50}{\normalcolor S50}, \hyperlink{S67}{\normalcolor S67}),
SMC (\hyperlink{S22}{\normalcolor S22}, \hyperlink{S41}{\normalcolor S41}),
LTSA (\hyperlink{S41}{\normalcolor S41}) \\
\cline{2-3}
& Scenario Modeling \& Language Support & 
VisualFML (\hyperlink{S22}{\normalcolor S22}), 
MSCs (\hyperlink{S26}{\normalcolor S26}), 
FisPro (\hyperlink{S22}{\normalcolor S22}), 
MOF (\hyperlink{S44}{\normalcolor S44}), 
DSLs (\hyperlink{S44}{\normalcolor S44}), 
CityGML (\hyperlink{S44}{\normalcolor S44}), 
MTSA (\hyperlink{S66}{\normalcolor S66}), 
Antlr (\hyperlink{S69}{\normalcolor S69}), 
WebGME (\hyperlink{S47}{\normalcolor S47}),
PLASMA (\hyperlink{S22}{\normalcolor S22}) \\
\hline
\multirow{4}{*}{\parbox{0.16\textwidth}{Semantic\\ Technologies}} 
& Ontology Development & 
Protégé (\hyperlink{S27}{\normalcolor S27}, \hyperlink{S31}{\normalcolor S31}, \hyperlink{S46}{\normalcolor S46}), 
RDF (\hyperlink{S04}{\normalcolor S04}, \hyperlink{S17}{\normalcolor S17}, \hyperlink{S34}{\normalcolor S34}), 
RDFS (\hyperlink{S04}{\normalcolor S04}), 
OWL (\hyperlink{S04}{\normalcolor S04}), 
Turtle (\hyperlink{S17}{\normalcolor S17}), 
SPARQL (\hyperlink{S17}{\normalcolor S17}),
SysML (\hyperlink{S74}{\normalcolor S74}) \\
\cline{2-3}
& Ontology Conversion & 
OWL2UML (\hyperlink{S27}{\normalcolor S27}), OWL2GRA (\hyperlink{S27}{\normalcolor S27}) \\
\cline{2-3}
& Data Exchange & 
JSON-LD (\hyperlink{S58}{\normalcolor S58}), XML (\hyperlink{S65}{\normalcolor S65}) \\
\cline{2-3}
& Centralized Supervision & 
CoDAMOS ontology (\hyperlink{S46}{\normalcolor S46}, \hyperlink{S54}{\normalcolor S54}) \\
\hline
\multirow{2}{*}{\parbox{0.16\textwidth}{Digital Twin \&\\ Containerization}} 
& Digital Twin & 
CONTACT Elements for integration (\hyperlink{S10}{\normalcolor S10}), Java EE Environment for deployment (\hyperlink{S60}{\normalcolor S60}) \\
\cline{2-3}
& Containerization & 
Docker (\hyperlink{S14}{\normalcolor S14}, \hyperlink{S45}{\normalcolor S45}, \hyperlink{S48}{\normalcolor S48}, \hyperlink{S58}{\normalcolor S58}), 
Kubernetes (\hyperlink{S14}{\normalcolor S14}, \hyperlink{S45}{\normalcolor S45}, \hyperlink{S48}{\normalcolor S48}), 
Docker Compose (\hyperlink{S64}{\normalcolor S64}) \\
\hline
\multirow{2}{*}{\parbox{0.16\textwidth}{Communication Middleware}} 
& Industrial Protocols & 
OPC UA (\hyperlink{S07}{\normalcolor S07}, \hyperlink{S12}{\normalcolor S12}, \hyperlink{S68}{\normalcolor S68}), 
FreeOpcUa/open62541 (\hyperlink{S68}{\normalcolor S68}), RAMI 4.0 (\hyperlink{S75}{\normalcolor S75})
 \\
\cline{2-3}
& Pub/Sub Systems & 
MQTT (\hyperlink{S10}{\normalcolor S10}, \hyperlink{S16}{\normalcolor S16}, \hyperlink{S36}{\normalcolor S36}, \hyperlink{S48}{\normalcolor S48}, \hyperlink{S52}{\normalcolor S52}, \hyperlink{S54}{\normalcolor S54}), 
Mosquitto MQTT (\hyperlink{S51}{\normalcolor S51}),
DDS (\hyperlink{S56}{\normalcolor S56}), 
RabbitMQ (\hyperlink{S64}{\normalcolor S64}), 
AMQP (\hyperlink{S17}{\normalcolor S17}), 
NSNotificationCenter (\hyperlink{S23}{\normalcolor S23}), 
MSMQ (\hyperlink{S23}{\normalcolor S23}),
EMQX (\hyperlink{S36}{\normalcolor S36}) \\
\hline
\multirow{2}{*}{\parbox{0.16\textwidth}{Extended Integration}} 
& Integration & 
HLA/RTI (Portico) (\hyperlink{S02}{\normalcolor S02}, \hyperlink{S03}{\normalcolor S03}, \hyperlink{S47}{\normalcolor S47}) \\
\cline{2-3}
& Service Discovery & 
Protocol Buffers (\hyperlink{S23}{\normalcolor S23}), 
DNS/DHCP (\hyperlink{S35}{\normalcolor S35}), 
MarkLogic Server (\hyperlink{S14}{\normalcolor S14}), 
THREDDS (\hyperlink{S14}{\normalcolor S14}) \\
\hline
\multirow{3}{*}{\parbox{0.16\textwidth}{Eclipse Ecosystem}} 
& Core Frameworks & 
Eclipse Arrowhead (\hyperlink{S05}{\normalcolor S05}, \hyperlink{S12}{\normalcolor S12}, \hyperlink{S19}{\normalcolor S19}, \hyperlink{S24}{\normalcolor S24}, \hyperlink{S52}{\normalcolor S52}, \hyperlink{S58}{\normalcolor S58}, \hyperlink{S63}{\normalcolor S63}),
Arrowhead Client Library (\hyperlink{S63}{\normalcolor S63}) \\
\cline{2-3}
& Device Management & 
Eclipse Hono (\hyperlink{S07}{\normalcolor S07}, \hyperlink{S12}{\normalcolor S12}, \hyperlink{S19}{\normalcolor S19}), 
Eclipse hawkBit (\hyperlink{S19}{\normalcolor S19}), 
Eclipse Ditto (\hyperlink{S19}{\normalcolor S19}), 
Eclipse Kura/Kapua (\hyperlink{S19}{\normalcolor S19}), 
Eclipse Vorto (\hyperlink{S19}{\normalcolor S19}), 
Eclipse 4diac\textsuperscript{TM} (\hyperlink{S12}{\normalcolor S12}), 
Eclipse Milo (\hyperlink{S07}{\normalcolor S07}) \\
\cline{2-3}
& Model-Driven Development & 
Fuzzy SosADL Studio (\hyperlink{S22}{\normalcolor S22}), 
Arrowhead Management Tool (\hyperlink{S21}{\normalcolor S21}), 
VIATRA (\hyperlink{S21}{\normalcolor S21}, \hyperlink{S44}{\normalcolor S44}), 
Sirius Framework (\hyperlink{S44}{\normalcolor S44}) \\
\hline
\multirow{2}{*}{\parbox{0.16\textwidth}{Architecture \&\\ Development}} 
& Architectural Modeling & 
UPDM (\hyperlink{S37}{\normalcolor S37}), 
SosADL (\hyperlink{S30}{\normalcolor S30}, \hyperlink{S57}{\normalcolor S57}), 
ArchSoS Language (\hyperlink{S42}{\normalcolor S42}) \\
\cline{2-3}
& Orchestration  & 
ESB (\hyperlink{S60}{S60})
 \\
\cline{2-3}
& Architecture Frameworks & 
NATO AF (\hyperlink{S74}{\normalcolor S74}), 
MAF (\hyperlink{S75}{\normalcolor S75}) \\
\hline
\multirow{2}{*}{\parbox{0.16\textwidth}{Robotic \& IoT Platforms}} 
& Robotic Platforms & 
LEGO MINDSTORMS EV3 (\hyperlink{S20}{\normalcolor S20}), 
ROS (\hyperlink{S65}{\normalcolor S65}, \hyperlink{S68}{\normalcolor S68}), 
POSIX (\hyperlink{S68}{\normalcolor S68}),
NVIDIA Isaac Sim (\hyperlink{S68}{\normalcolor S68}) \\
\cline{2-3}
& IoT Devices & 
Raspberry Pi-based systems (\hyperlink{S70}{\normalcolor S70}, \hyperlink{S68}{\normalcolor S68}) \\
\hline
\multirow{3}{*}{\parbox{0.16\textwidth}{Data Management, Cloud \&\\ Infrastructure}} 
& Data Flow & 
SPIDER (\hyperlink{S59}{\normalcolor S59})\\
\cline{2-3}
& Data Optimization & 
ElasticSearch/OpenSearch (\hyperlink{S14}{\normalcolor S14}) \\
\cline{2-3}
& Cloud Provisioning & 
MITRE Symphony (\hyperlink{S36}{\normalcolor S36}), 
Terraform (\hyperlink{S36}{\normalcolor S36}), 
Ansible (\hyperlink{S36}{\normalcolor S36}) \\
\hline
\multirow{3}{*}{\parbox{0.16\textwidth}{Workflow, Low-Code \&\\ Testing Tools}} 
& Workflow Execution & 
Node-RED (\hyperlink{S16}{\normalcolor S16}) \\
\cline{2-3}
& Low-Code \& Integration & 
MicrographQL Tool (\hyperlink{S06}{\normalcolor S06}), 
StringTemplate (\hyperlink{S18}{\normalcolor S18}),
DIME (\hyperlink{S56}{\normalcolor S56}) \\
\cline{2-3}
& Testing \& Modeling & 
MATLAB Fuzzy Logic Toolbox (\hyperlink{S22}{\normalcolor S22}), 
AspectJ (\hyperlink{S26}{\normalcolor S26}), 
Prolog Rule Engines (\hyperlink{S34}{\normalcolor S34}), 
mKAOS (\hyperlink{S15}{\normalcolor S15}), 
W3C WoT (\hyperlink{S58}{\normalcolor S58}), 
IoT Broker (\hyperlink{S60}{\normalcolor S60}), 
Orthus Blockchain (\hyperlink{S64}{\normalcolor S64}), 
Apache JMeter (\hyperlink{S64}{\normalcolor S64}) \\
\hline
\multirow{1}{*}{\parbox{0.16\textwidth}{AI/ML}} 
& AI/ML Frameworks & 
Fine-tuned BERT (\hyperlink{S32}{\normalcolor S32}), 
AI/ML for data processing (\hyperlink{S39}{\normalcolor S39}), 
Machine learning for optimization (\hyperlink{S40}{\normalcolor S40}), 
Deep RL (\hyperlink{S72}{\normalcolor S72}), 
Neural Networks (\hyperlink{S11}{\normalcolor S11}) \\
\bottomrule
\end{tabular}
\end{table*}

\textit{Simulation \& Scenario Testing} tools dominate the ecosystem, ranging from co-simulation platforms like Mosaik (S01, S18, S47) to high-fidelity simulators such as Gazebo (S40, S51) and Unity (S40, S48), which validate dynamic behaviors in domains like drone control (VENTOS; S33) and urban mobility (SUMO; S48). These are complemented by \textit{Formal Modeling \& Verification} tools like PRISM (probabilistic verification; S33, S41) and SysML (model-driven design; S21, S37), which ensure rigor in interoperability and safety.

\textit{Semantic Technologies}---such as Protégé (S27, S31, S46) and RDF-based approaches (S04, S17, S34)---alongside communication middleware like Open Platform Communications-Unified Architecture (OPC UA)  OPC UA (S07, S12, S68) and Message Queuing Telemetry Transport (MQTT) (S10, S16, S48), reflect efforts to standardize cross-domain data exchange and promote interoperability.

The \textit{Eclipse Ecosystem} stands out as a major contributor, with integrated frameworks and tools including EAF, Hono, HawkBit, Ditto, Kapua, and Corto appearing across 11 studies (S05, S07, S12, S19, S21, S22, S24, S44, S52, S58, S63). Its widespread adoption highlights the growing demand for cohesive, extensible solutions for industrial SoS development.

Other technologies support deployment and interaction across physical and digital layers. \textit{Containerization} platforms like Docker and Kubernetes (S14, S45, S48) enable modular and agile deployment, while \textit{Digital Twin} frameworks (e.g., CONTACT Elements; S10, S60) facilitate runtime monitoring and adaptation. Similarly, \textit{Robotic and IoT Platforms} such as ROS (S65, S68) help bridge hardware-software integration in real-world settings.

Well-established tools like EAF, MQTT, ROS, and Docker reflect maturity for real-world deployment. In contrast, AI or machine learning (ML) techniques---including fine-tuned bidirectional encoder representations from transformers (BERT) (S32), Deep RL (DRL) (S72), and Neural Networks (S11)---are underutilized, signaling their emerging status rather than integration maturity. Similarly, blockchain (Orthus; S64) and low-code tools (Node-RED; S16) lack cohesive adoption, operating in isolation rather than as part of unified toolchains.

Despite the rich ecosystem, tool integration remains fragmented. 
While platforms like EAF (S05, S63) enable service orchestration and DIME (S56) supports low-code integration, few studies demonstrate complete toolchains spanning from high-level modeling to simulation to runtime adaptation (e.g., NVIDIA Omniverse; S68). 
This absence highlights a gap in holistic SoSE environments capable of supporting the whole system lifecycle.

\subsection{Evaluation (RQ\textsubscript{4})}
\label{sec:evaluation}

This section explores how studies addressing runtime composition in SoSs evaluate their proposed solutions. We present our findings organized into two complementary aspects: 
1)  the methodologies employed to assess solutions (Section~\ref{sec:evaluation-methods}), 
and
2) the application domains and case studies contextualizing these evaluations(Section~\ref{sec:application-domains-and-case-studies}).

\subsubsection{Evaluation Methods}
\label{sec:evaluation-methods}

Table~\ref{tbl:evaluation_methods} provides the taxonomy of the evaluation methods employed in the proposed solutions. 
These themes and their respective subthemes are elaborated in the remainder of this section.

\begin{table*}[t]
\footnotesize
\caption{Taxonomy of evaluation methods.}
\label{tbl:evaluation_methods}
\begin{tabular}{p{0.12\textwidth}p{0.18\textwidth}p{0.65\textwidth}}
\rowcolor[rgb]{0.90,0.90,0.90}
\hline
\textbf{Methods} & \textbf{Subthemes} & \textbf{Evaluation methods with corresponding study IDs} \\
\hline
\multirow{25}{*}{\parbox{0.12\textwidth}{Simulation}} 
& \multirow{4}{*}{\parbox{0.18\textwidth}{Co-Simulation}} 
& $\bullet$ Co-simulation of modular production units in industrial automation using Mosaik (\hyperlink{S01}{\normalcolor S01}) \\
& & $\bullet$ Co-simulation of EV fleet ecosystems for energy/charging/infrastructure load analysis (\hyperlink{S48}{\normalcolor S48}) \\
& & $\bullet$ Simulation of C3 federation (\hyperlink{S03}{\normalcolor S03}) \\
& & $\bullet$ Simulation for varying interoperability parameters in SAR missions (\hyperlink{S49}{\normalcolor S49}) \\
\cline{2-3}
& \multirow{3}{*}{\parbox{0.18\textwidth}{Scenario-Based \\ Simulation}} 
& $\bullet$ For validating reconfiguration strategies and SoS agility (\hyperlink{S11}{\normalcolor S11}, \hyperlink{S25}{\normalcolor S25}, \hyperlink{S34}{\normalcolor S34}, \hyperlink{S37}{\normalcolor S37}) \\
& & $\bullet$ To generate scenarios for evaluating performance (\hyperlink{S42}{\normalcolor S42}, \hyperlink{S44}{\normalcolor S44}, \hyperlink{S41}{\normalcolor S41}, \hyperlink{S35}{\normalcolor S35}, \hyperlink{S28}{\normalcolor S28}, \hyperlink{S39}{\normalcolor S39}, \hyperlink{S17}{\normalcolor S17}, \hyperlink{S77}{\normalcolor S77}, \hyperlink{S66}{\normalcolor S66}, \hyperlink{S33}{\normalcolor S33}, \hyperlink{S58}{\normalcolor S58}) \\
& & $\bullet$ Scenario-based simulation for adaptation strategy generation (\hyperlink{S26}{\normalcolor S26}, \hyperlink{S73}{\normalcolor S73}) \\
\cline{2-3}
& \multirow{2}{*}{\parbox{0.18\textwidth}{Agent-Based Simulation}} 
& $\bullet$ Iterative design and performance analysis using agent-based models (\hyperlink{S62}{\normalcolor S62}) \\
& & $\bullet$ Multi-actor strategic behavior analysis using game-theoretic value modeling (\hyperlink{S04}{\normalcolor S04}) \\
\cline{2-3}
& \multirow{5}{*}{\parbox{0.18\textwidth}{Discrete-Event and \\ Numerical Simulation}} 
& $\bullet$ Evaluating performance of autonomous systems under adversarial conditions (\hyperlink{S29}{\normalcolor S29}) \\
& & $\bullet$ OMNeT++ simulation for Mesos cluster scaling (\hyperlink{S46}{\normalcolor S46}) \\
& & $\bullet$ Formal simulation of fuzzy SoS architectures using DEVS-Suite/PLASMA (\hyperlink{S22}{\normalcolor S22}) \\
& & $\bullet$ DEVS models in MS4ME environment with ASAS method (\hyperlink{S57}{\normalcolor S57}) \\
& & $\bullet$ Maude-based rewriting logic simulation for resource allocation (\hyperlink{S67}{\normalcolor S67}) \\
\cline{2-3}
& \multirow{8}{*}{\parbox{0.18\textwidth}{Autonomous Systems \\ Simulation (UxV/UAV/UGV)}} 
& $\bullet$ NetLogo/ESimP in SEAT execution environment (\hyperlink{S08}{\normalcolor S08}) \\
& & $\bullet$ Gazebo/DroneResponse for sUAS (\hyperlink{S51}{\normalcolor S51}) \\
& & $\bullet$ JADE-based UVF coordination (\hyperlink{S80}{\normalcolor S80}) \\
& & $\bullet$ Monte Carlo simulation for UAV swarm reliability and failure response (\hyperlink{S53}{\normalcolor S53}) \\
& & $\bullet$ Factory simulation of autonomous transport vehicles for safety evaluation (\hyperlink{S61}{\normalcolor S61}) \\
& & $\bullet$ Multi-swarm reconfiguration under attack scenarios (\hyperlink{S71}{\normalcolor S71}) \\
& & $\bullet$ Simulating UWSoS attack scenarios with varying levels of disruption (\hyperlink{S72}{\normalcolor S72}) \\
& & $\bullet$ Hybrid cloud simulation for mission scenarios (\hyperlink{S36}{\normalcolor S36}) \\
\hline
\multirow{4}{*}{\parbox{0.1\textwidth}{Implementation-Based Evaluation}} 
& \multirow{2}{*}{Testbeds} 
& $\bullet$ Hardware-in-the-Loop Testing (\hyperlink{S02}{\normalcolor S02}, \hyperlink{S20}{\normalcolor S20}, \hyperlink{S07}{\normalcolor S07}, \hyperlink{S56}{\normalcolor S56}, \hyperlink{S46}{\normalcolor S46}, \hyperlink{S65}{\normalcolor S65}, \hyperlink{S68}{\normalcolor S68}) \\
& & $\bullet$ Benchmarking accuracy/scalability of tools/algorithms (\hyperlink{S32}{\normalcolor S32}) \\
\cline{2-3}

& Proof-of-Concepts 
& $\bullet$ Validation of technical feasibility  (\hyperlink{S06}{\normalcolor S06}, \hyperlink{S13}{\normalcolor S13}, \hyperlink{S21}{\normalcolor S21}, \hyperlink{S27}{\normalcolor S27}, \hyperlink{S31}{\normalcolor S31}, \hyperlink{S47}{\normalcolor S47}, \hyperlink{S52}{\normalcolor S52}, \hyperlink{S55}{\normalcolor S55}, \hyperlink{S70}{\normalcolor S70}, \hyperlink{S66}{\normalcolor S66}) \\
\cline{2-3}

& Prototype Implementations
& $\bullet$ Functional implementation for practical validation (\hyperlink{S12}{\normalcolor S12}, \hyperlink{S14}{\normalcolor S14}, \hyperlink{S16}{\normalcolor S16}, \hyperlink{S23}{\normalcolor S23}, \hyperlink{S59}{\normalcolor S59}, \hyperlink{S63}{\normalcolor S63}, \hyperlink{S60}{\normalcolor S60}, \hyperlink{S64}{\normalcolor S64}, \hyperlink{S69}{\normalcolor S69}, \hyperlink{S24}{\normalcolor S24}, \hyperlink{S17}{\normalcolor S17}, \hyperlink{S30}{\normalcolor S30}) \\
\hline
\multirow{9}{*}{\parbox{0.12\textwidth}{Human-Centered Evaluation}} 
& \multirow{3}{*}{\parbox{0.18\textwidth}{User Studies}} 
& $\bullet$ Interviews with non-specialists (\hyperlink{S74}{\normalcolor S74}, \hyperlink{S79}{\normalcolor S79}) \\
& & $\bullet$ User experiments (\hyperlink{S43}{\normalcolor S43}) \\
& & $\bullet$ Developer user study (\hyperlink{S54}{\normalcolor S54}) \\
\cline{2-3}
& \multirow{3}{*}{\parbox{0.18\textwidth}{Expert Assessment}} 
& $\bullet$ Semi-structured interviews (\hyperlink{S15}{\normalcolor S15}) \\
& & $\bullet$ Architecture Trade-off Analysis Method (ATAM) (\hyperlink{S34}{\normalcolor S34}) \\
& & $\bullet$ Expert questionnaire evaluation (\hyperlink{S76}{\normalcolor S76}) \\
\cline{2-3}
& \multirow{2}{*}{\parbox{0.18\textwidth}{Collaborative Evaluation}} 
& $\bullet$ Discussion-based evaluation (\hyperlink{S47}{\normalcolor S47}) \\
& & $\bullet$ Multi-stakeholder workshops (\hyperlink{S74}{\normalcolor S74}) \\
\hline
\multirow{3}{*}{\parbox{0.12\textwidth}{Formal \& \\ Static Analysis}} 
& \multirow{2}{*}{\parbox{0.18\textwidth}{Formal Analysis}} 
& $\bullet$ Model checking (\hyperlink{S42}{\normalcolor S42}, \hyperlink{S50}{\normalcolor S50}) \\
& & $\bullet$ Formally verified implementations (\hyperlink{S33}{\normalcolor S33}, \hyperlink{S55}{\normalcolor S55}) \\
\cline{2-3}
& \multirow{1}{*}{\parbox{0.18\textwidth}{Static Analysis}} 
& $\bullet$ Graph-based methods for resilience, robustness, and inter-agent communication analysis (\hyperlink{S62}{\normalcolor S62}) \\
\bottomrule
\end{tabular}
\end{table*}

\paragraph{Simulation-Based Evaluation} 
methods dominate the landscape, likely due to their ability to model complex interactions without incurring the substantial resource demands of full-scale implementations.
Among these methods, \textit{Scenario-Based Simulation} is the most prevalent category, with researchers generating various scenarios to evaluate performance, develop adaptation strategies, and validate reconfiguration strategies.
In addition, \textit{Autonomous Systems Simulation} forms a distinct cluster (8 studies), reflecting the growing importance of autonomous systems in SoS contexts. 
These ranged from unmanned aerial vehicle (UAV) swarm reliability assessments using Monte Carlo methods (S53) to multi-swarm reconfigurations under stack scenarios (S71).
\textit{Co-Simulation} enables the integration of different simulators to study interactions between CSs in domains such as industrial automation (S01) and electric vehicle (EV) fleet management (S48).
\textit{Agent-Based Simulation} models CSs as autonomous agents, enabling iterative design and strategic analysis of emergent behaviors (S62, S04).
\textit{Discrete-Event Simulations}  capture SoS  responses under specific events or adverse conditions (S29), providing insights into resilience and adaptability.

\paragraph{Implementation-Based Evaluation} 
is the second most prevalent approach, emphasizing real-world applicability and feasibility. 
Most studies developed prototypes (12 studies) and proof-of-concepts (10 studies) to validate the technical feasibility of their solutions. 
Hardware-in-the-Loop testing was particularly prominent within \textit{Testbed} implementations, enabling evaluation under conditions closely approximating real-world operational environments.

\paragraph{Human-Centered Evaluation}
although less frequently employed, is critical for evaluating the usability and stakeholder acceptance of SoS solutions.
These approaches include user studies with non-specialists (S74, S79), developer-focused evaluations (S54), expert assessments via techniques like the Architecture Trade-off Analysis Method (S34), and structured interviews (S15), and collaborative evaluation techniques including multi-stakeholder workshops (S74).

\paragraph{Formal and Static Analysis} 
provide rigorous verification capabilities through \textit{model checking} (S42, S50) and formally \textit{verified implementations} (S33, S55). 
These methods appeared in relatively few studies, suggesting potential for expanded application in future research.

\subsubsection{Application Domains and Case Studies}
\label{sec:application-domains-and-case-studies}

% Many studies  apply specific evaluation methods and situate their validations within well-defined application domains and real-world case scenarios.
Most studies situate their evaluation activities within well-defined application domains and real-world case scenarios.
This contextual grounding is essential for assessing the practical relevance and transferability of the proposed solutions.
Therefore, we systematically mapped the case studies and application domains across the selected studies.

\begin{figure*}[t]
    % edit link: https://app.mindmup.com/map/_free/2025/03/6465e59000dc11f0bac8b18079d6db62
    \centering
    \includegraphics[width=0.9\linewidth]{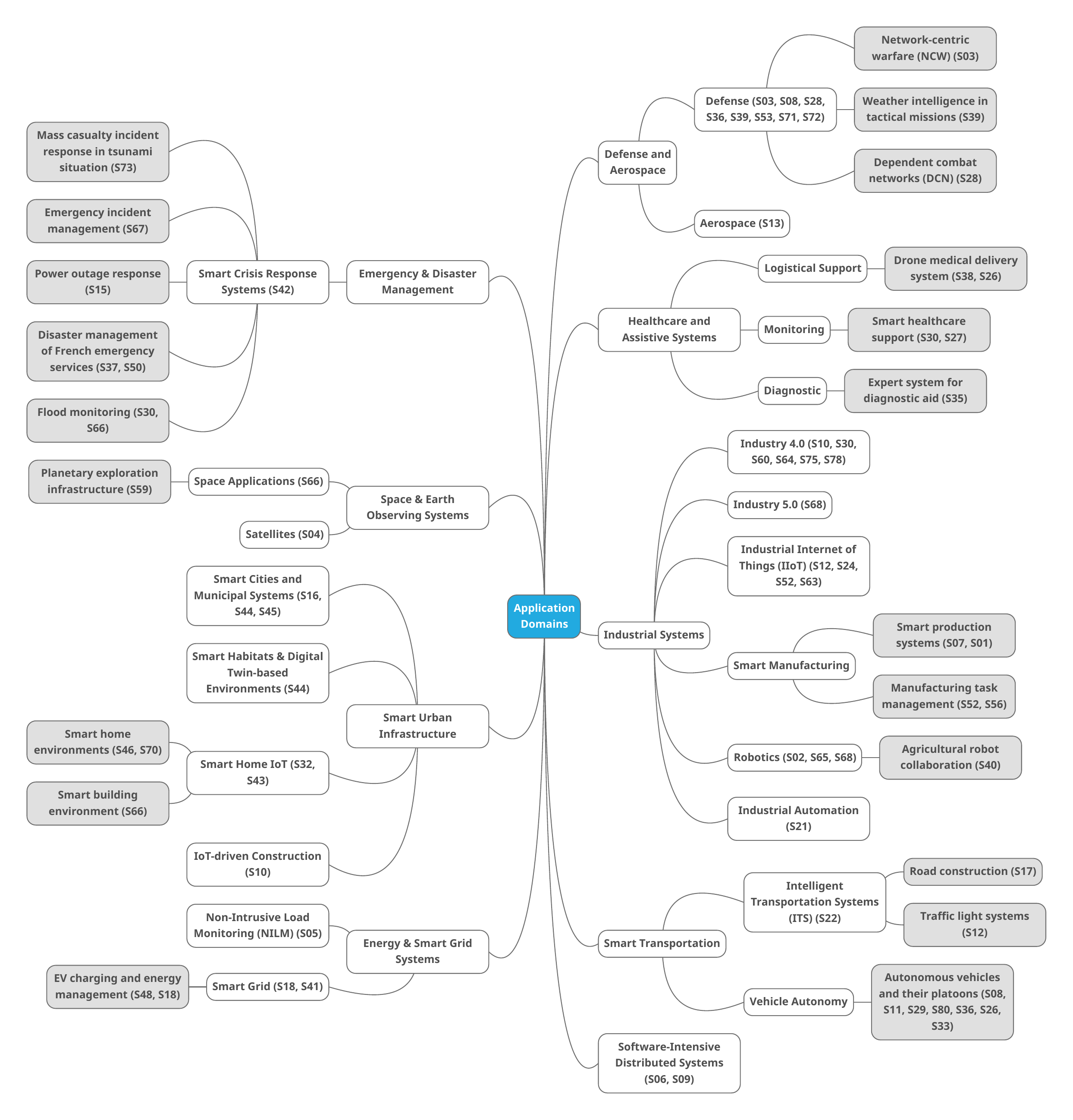}
    \caption{Application domains identified in the selected studies. Unshaded nodes represent domains and subdomains, while shaded nodes indicate case studies.}
    \label{fig:application-domains-and-case-studies}
\end{figure*}

As shown in Fig.~\ref{fig:application-domains-and-case-studies}, our analysis reveals a diverse ecosystem of domains where SoS approaches are deployed,  from industrial automation to emergency response. 
In this figure, shaded nodes represent  case studies, while unshaded nodes indicate application domains.
Notably, some application domains lack associated case studies, such as Satellites (S04), Aerospace (S13), and Industry 5.0 (S68), as these studies mention a domain but do not evaluate their approach with a case study. Furthermore, some studies assess their approach using multiple case studies (S66).
Among these domains, industry, defense, and disaster management are the most prominent application areas, highlighting their critical role in addressing complex, real-world challenges.

\section{Discussion}
\label{sec:discussion}

This section synthesizes the results, explores research gaps and potential future directions, and discusses the implications for  research and practice.

\subsection{Key Findings and Interpretations}
This section presents key findings of the results, their interpretation, and critical analysis, and situates them within the existing literature.

\subsubsection{Fragmented Research Communities}

The fragmented distribution of research across diverse venues  (Section~\ref{sec:venue-distribution}), with limited overlap, underscores both the challenges and opportunities inherent in the research subject. 
While this fragmentation reflects an emergent, interdisciplinary field shaped by diverse perspectives and methodologies, it  risks redundant efforts and stalled progress due to limited cross-community collaboration. 
By systematically connecting these siloed discourses, the SLR  addresses the immediate need for synthesis to mitigate fragmentation and highlights the potential for convergence, fostering interdisciplinary dialogue and accelerating innovation in this evolving domain.

\subsubsection{The Challenge Landscape of Runtime Composition in SoSs}

The challenges to enabling runtime composition in SoSs (RQ\textsubscript{1}) range from technical interoperability concerns to architectural governance issues.

\paragraph{Multi-dimensional Heterogeneity.}
The heterogeneity in SoS extends well beyond technical incompatibility: it also encompasses proprietary restrictions and operational coupling that introduce systemic rigidity, posing a significant barrier that requires interdisciplinary solutions.
We discovered (Section~\ref{sec:Heterogeneity of Constituent Systems}) that reconciling differences in physical devices, data formats, communication protocols, and development methodologies is crucial for ensuring seamless integration, strong performance, and flexible operations.
We believe that this issue of heterogeneity is one of the key factors that necessitate moving beyond traditional approaches, as highlighted by \citep{9593274}.

\paragraph{The Modeling-Reality Gap.} Our analysis reveals a persistent gap between design-time models and real-world SoS behavior, particularly in capturing emergent properties (Section~\ref{sec:Modeling and Analysis}) and multi-domain interactions. This  gap aligns with the difficulty in managing architectural emergence and integration~\citep{9593274}. Our taxonomy contextualizes these challenges regarding runtime composition and reveals why traditional verification and validation remain problematic, even with advanced models. The exponential increase in verification costs during frequent reconfigurations (S33, S41) highlights the inadequacy of static techniques for dynamic SoSs. These findings underscore the need for unified, lightweight frameworks that enable continuous validation and close the gap between modeling assumptions and runtime realities.

\paragraph{The Autonomy-Coordination Paradox.}
Evident in conflicts between decentralized control strategies (S50) and mission coherence requirements (S42), this reflects a core tension in \textit{System Orchestration} theme (Section~\ref{sec:system_orchestration}). 
Solutions like hybrid architectures (S69) and Discrete Controller Synthesis (S66) attempt to balance these priorities but lack standardized evaluation metrics.
This paradox appears across multiple subthemes such as \textit{Dynamic Discovery \& Composition}, \textit{Control Strategies}, and \textit{Mission Coherence \& Operational Continuity}. 
These limitations become especially problematic under volatile environmental conditions or component-level faults, where miscoordination can threaten operational continuity or trigger cascading failures. Effective solutions must embrace this paradox by developing adaptive strategies that maintain mission coherence while balancing autonomy, resilience, and context-aware coordination.

\subsubsection{Evolution of Solution Approaches}
Our analysis of solutions (Section~\ref{sec:solutions}) reveals several notable trends in addressing the identified challenges. While diverse solutions exist, the landscape is fragmented, with SOA being the noteworthy exception. Additionally, we observe a fundamental tension between achieving domain-specific precision and enabling cross-domain scalability. 

\paragraph{The Shift Toward Adaptive Architectures.}
Table~\ref{tbl:solutions} demonstrates an evolutionary trajectory from static to dynamic architectural approaches. Early solutions focused on standardized interfaces and protocols (S04, S45), while more recent approaches emphasize adaptive architectures (S69, S22) and dynamic reconfiguration capabilities (S80, S27). This evolution reflects growing recognition that static interoperability solutions cannot adequately address the inherent dynamism of modern SoS environments~\citep{10.1145/3659098}. 

The emergence of hybrid approaches that combine formal methods with dynamic adaptation mechanisms (S69, S22, S55, S42) is particularly noteworthy.
These hybrid approaches suggest a maturing field  moving beyond the false dichotomy of formal rigor versus adaptive flexibility.

This shift aligns with observations by \cite{9593274}, who noted the diminishing relevance of traditional architectural approaches where systems autonomously self-organize without centralized control. Our findings extend this observation by identifying  emerging architectural patterns  to address these dynamic contexts, particularly in runtime composition scenarios.

\paragraph{Convergence Around Service-Oriented Paradigms.}
The convergence around SOA approaches, with particular emphasis on microservices (S06, S60) and standardized frameworks like the EAF (S05, S16, S52, S12, S21, S58, S19, S63), suggests an emerging consensus around their suitability for SoS integration challenges.

These findings align with \cite{10.5555/3492252.3492254}, who identified SOA and MSA among key architectural patterns for addressing SoS requirements. However, our analysis goes deeper by revealing  the prevalence of these patterns and their specific implementations and adaptations for runtime composition contexts.

This convergence may also indicate a consolidation phase in the field, where researchers  build upon established architectural patterns rather than propose fundamentally new approaches. However, it also raises concerns about potential limitations in addressing challenges  outside the service-oriented paradigm, such as physical integration and emergent behaviors in CPS contexts.

\paragraph{The Rise of Semantic Approaches.}
Semantic technologies, particularly ontology-based approaches (S32, S46, S74), address dynamic discovery and interoperability in SoS contexts by enabling semantic understanding of capabilities and constraints. 
% These technologies reflect a shift beyond syntactic interoperability, aligning with \cite{s24092921}'s emphasis on ontology alignment and linked-data inference. 
These technologies reflect a shift beyond syntactic interoperability, aligning with emphasis on ontology alignment and linked-data inference~\citep{s24092921}.
Our analysis highlights concrete implementations in SoS, such as integrating ontologies with holonic architectures (S46) and service discovery (S32), demonstrating the complementary use of semantic and architectural solutions.

% These findings extend \cite{NAUDET2023695}'s cognitive interoperability framework for CPS, providing practical examples of semantic integration in SoS environments.
These findings extend the cognitive interoperability framework for CPS proposed by \cite{NAUDET2023695}, providing practical examples of semantic integration in SoS environments.

\subsubsection{The Tooling Ecosystem}
Our analysis of the tooling ecosystem (RQ\textsubscript{3}) reveals both areas of maturity and  gaps that may impede the practical implementation of runtime composition in SoS.

\paragraph{Simulation-Implementation Gap.}
While our results show rich ecosystems in both simulation tools (e.g., Mosaik, Gazebo) and implementation frameworks (e.g., EAF, Docker), the fragmentation of these tools (most operating in isolation) means there is a notable lack of integrated toolchains bridging the simulation and implementation phases (with NVIDIA Omniverse in S68 as a rare exception). This disconnect between simulation‑validated approaches and practical deployments may explain why many promising solutions fail to transition to practice. 

\paragraph{The Eclipse Ecosystem.}

This ecosystem emerged as a significant contributor, highlighting the demand for cohesive, extensible solutions and representing a positive trend toward greater interoperability and knowledge sharing in the field. 
% However, the review identified a gap in holistic SoSE environments capable of supporting the whole system lifecycle. 

\subsubsection{Evaluation Methods}
Our analysis of evaluation methods (RQ\textsubscript{4}) reveals several patterns that have implications for assessing the effectiveness of the proposed  solutions.
This analysis aligns with \cite{10.1145/3519020}, who focused on evaluating SoS architectures. Our study builds on this by broadly examining  SoSE strategies, including a detailed taxonomy of evaluation methods and application domains.

\paragraph{Simulation Dominance: Strength or Limitation?}
The dominance of simulation-based evaluation methods, particularly scenario-based simulations, aligns with findings from \cite{9078791}, who investigated agent-based modeling  to simulate SoS. 
Our analysis extends their work by identifying a broader range of simulation approaches beyond agent-based modeling and critically examining the limitations of simulation-centric evaluation in  runtime composition. 
However, this reliance on simulation may inadvertently bias research toward solutions that perform well in simulated environments but may not address the full complexity of real-world SoS challenges with emergent behaviors and unanticipated interactions. 

\paragraph{Domain Diversity: Strength in Breadth, Weakness in Depth.}

The diversity of application domains identified in our analysis---from industrial automation to emergency response---demonstrates the broad relevance of runtime composition across different SoS contexts.
However, this diversity also results in fragmented evaluation methods that make comparing solutions across domains or developing domain-specific best practices difficult.

This fragmentation is particularly evident in the literature's limited representation of specific SoS types---directed and collaborative. Most studies (16) refer to general SoS without specifying a particular kind, suggesting insufficient attention to how different SoS contexts might require different runtime composition approaches.
The absence of these SoS types suggests an opportunity for future research to address them explicitly.

This fragmentation is underscored by the limited attention given to specific SoS types, particularly directed and collaborative systems, which are not explicitly addressed in  reviewed studies. Most studies 
refer only to general SoS without specifying a type, suggesting a lack of consideration for how varying SoS contexts may require distinct runtime composition strategies. This fragmentation highlights a promising direction for future research to  investigate and characterize underrepresented SoS types more explicitly..

\paragraph{Insufficient Attention to Scalability.}

Many studies demonstrate solutions in controlled environments with limited CSs, raising questions about scalability to large-scale SoS with hundreds or thousands of CSs. This limitation is particularly concerning given the growing scale and complexity of real-world SoS deployments.

\subsection{Research Gaps and Future Directions}
Based on our analysis of results, we identify the following research gaps and directions for future work.

\subsubsection{Bridging Technical and Socio-Technical Dimensions}
While the literature provides rich coverage of technical challenges and solutions for runtime composition, there is insufficient attention to the socio-technical dimensions that often determine practical success. This gap aligns with findings from \cite{10.1145/3659098}, who identified a significant shift in interoperability research from purely technical concerns to socio-technical issues. 
% Their tertiary study highlighted that interoperability now extends beyond software systems and requires multidisciplinary approaches---a need echoed by the organizational and human factor gaps identified in RQ1---that our findings confirm are still lacking in SoS contexts.
Their tertiary study highlighted that interoperability now extends beyond software systems and requires multidisciplinary approaches. This need is echoed by the organizational and human factor gaps identified in RQ\textsubscript{1}, which our findings show are still lacking in SoS contexts.

Future research should develop frameworks that explicitly address how organizational, human, commercial, and regulatory factors interact with technical solutions in SoS environments. Particularly promising are approaches that combine technical standards with governance frameworks, incentive mechanisms, and business models that encourage interoperability across organizational boundaries. Such  holistic approaches are essential for addressing the proprietary constraints and organizational coupling identified in our challenge taxonomy.

\subsubsection{AI-Enabled Adaptation and Orchestration}

The limited penetration of AI and machine learning (ML) techniques in current solutions represents  future research opportunities. AI approaches could address several  challenges identified in our taxonomy, including emergence management, dynamic discovery, and adaptive control strategies.

This gap aligns with findings from \cite{s24092921}, highlighting the underrepresentation of ML techniques in SoS research. Our analysis confirms this gap while providing specific examples of where AI/ML could be most effectively applied in runtime composition contexts.

Particularly promising are RL approaches for developing adaptation policies in uncertain environments (S73) and graph neural networks for inferring resilience strategies (S72). These approaches can  address challenges identified by \cite{10.1145/3487921} regarding uncertainty in self-adaptive systems, particularly in managing unanticipated changes. Future research should explore how these techniques can be integrated into comprehensive frameworks that address the \textit{autonomy-coordination paradox} while maintaining safety and predictability.

\subsubsection{The Benchmarking Problem}
A  methodological gap emerges from our analysis of evaluation methods. Current research predominantly validates solutions through isolated simulations or proof-of-concept implementations based on scenario-specific assumptions (Table~\ref{tbl:evaluation_methods}). 
This fragmentation creates two fundamental problems: evaluation metrics remain inconsistently defined across studies, and proposed solutions are rarely benchmarked against competing approaches.

While these isolated evaluations provide valuable insights into individual solutions, they prevent rigorous comparative assessment. This is unlike the ML domain, where standardized protocols, benchmarking datasets, and unified metrics accelerate progress by requiring new methods to demonstrate measurable improvements over existing baselines, fostering reproducibility and cumulative advancement.

The absence of standard benchmarks in SoS runtime composition research  impedes cross-approach comparison and slows innovation momentum. Most studies compare against only self-developed baselines, except a few  utilize common reference architectures. This absence undermines claims of generalizability and hinders progress, particularly for challenges requiring cross-organizational collaboration.

Future research should prioritize developing  benchmarks that reflect real-world SoS constraints, including (1) standardized metrics that capture both technical performance  and socio-technical factors and (2) shared platforms supporting cross-domain evaluations. Such standardization would  facilitate more robust evaluation and foster the collaboration and incremental improvement that have driven rapid advancement in other technical fields.

\subsection{Practitioner Recommendations}

Our findings have several implications for practitioners working on SoS development and integration.

\subsubsection{Architectural Choices: Service-Orientation as Foundation}

The convergence around service-oriented architectural approaches, particularly the EAF, suggests that practitioners  consider these paradigms  foundational for new SoS developments. The widespread adoption across multiple studies indicates a level of maturity and community support that can reduce implementation risks.

This recommendation aligns with the catalog of design patterns presented by \cite{9812681}, who identified service-oriented approaches as one of the key patterns for SoS. Our findings extend this work by providing specific evidence of successful implementations and highlighting the EAF as a  promising standardization effort.

However, practitioners should know the limitations of purely service-oriented approaches, particularly for systems with significant physical components or real-time constraints. In such cases, complementary approaches, e.g., digital twins (S40, S61), may be necessary to address the complete integration challenges. This consideration is essential in light of findings from \cite{10178527} regarding the challenges of digital twins in SoS contexts, highlighting the need for a deeper understanding of their concepts, principles, costs, and benefits.

\subsubsection{Tool Selection Strategy: Ecosystem Over Individual Tools}

Practitioners should prioritize coherent tool ecosystems over individual tool capabilities when selecting technology. The prominence of integrated ecosystems like Eclipse in the literature indicates the importance of tool interoperability and community support for successful SoS implementation.

When selecting tools, practitioners should evaluate  current capabilities and ecosystem vitality, standardization efforts, and integration with complementary tools spanning the entire SoS lifecycle---from design to deployment and operation. 
This ecosystem-focused approach can help avoid the fragmentation challenges identified in our analysis and should be complemented by consideration of organizational and human factors that influence successful adoption.

\section{Threats to Validity}
\label{sec:threats-to-validity}

\updated{
SLRs inevitably face certain limitations and risks that may impact the trustworthiness of their findings. 
This section outlines the key threats to validity inherent to the study's context and design, and describes the strategies employed to mitigate their influence. 
While many threats were addressed through rigorous methods, some residual limitations remain unavoidable and are transparently acknowledged to support critical appraisal and future replication efforts.}

\subsection{External Validity}
\label{sec:external_validity}

\updated{External validity---the extent to which our findings can be generalized beyond the specific studies and contexts included---can be jeopardized if the search process misses relevant work. 
To mitigate this risk, we 
(1) performed comprehensive searches across six leading digital libraries (chosen for their broad coverage, quality content, and prevalent use in SLR practice), 
(2) constructed a robust search string incorporating multiple keyword combinations drawn from established SoS terminology (Section~\ref{sec:search-strategy}), and 
(3) complemented the automated search with backward and forward snowballing.  }

\updated{
Consistent with Wohlin's guidelines~\citep{wohlin2014guidelines}, snowballing was performed using the 77 primary studies retained after title/abstract screening, full-text screening, and quality assessment, ensuring a reliable start set. This avoided the infeasibility of applying snowballing to thousands of raw search hits while still maximizing coverage. Following \cite{kitchenham2015evidence}, the same inclusion and exclusion criteria were systematically applied, and candidate studies meeting these criteria were added to the corpus. 
While primarily a safeguard for external validity, snowballing also contributes to internal validity (Section~\ref{sec:internal_validity}) by reducing the risk of omitting relevant studies due to search string limitations.  
% To further guard against omissions, we cross-checked our dataset against the reference lists of key primary studies and recent domain surveys (Section~\ref{sec:related-work}). Any study identified through these sources that met our inclusion criteria was added to the corpus. 
This two-pronged approach increases confidence in the completeness of our dataset.  
}

\updated{We acknowledge, however, that the five-year publication window (2019--2024) may have excluded earlier but still relevant foundational studies. This constraint was necessary to ensure the feasibility of the review, given the high volume of recent publications in the field. As \cite{kitchenham2015evidence} note, such scoping decisions are acceptable when they make in-depth synthesis manageable. Future reviews could expand the temporal scope to explore longitudinal trends.  

Finally, including low-quality or tangentially relevant work can bias conclusions. We applied the established SLR quality‐assessment checklist to every candidate study, filtering out those failing to meet our predefined thresholds of methodological rigor and venue reputation. Only studies that satisfied all quality criteria were retained.
}

\subsection{Internal Validity}
\label{sec:internal_validity}

Internal validity refers to the soundness and consistency of our thematic classifications and interpretations for each RQ.
A\textsubscript{1} and A\textsubscript{3} independently drafted initial classification schemes in parallel to generate candidate taxonomies. The A\textsubscript{1}  then consolidated these drafts into a single taxonomy, refining definitions and resolving  discrepancies through full‐text review and iterative discussion. 

While we did not perform a formal inter‐rater reliability statistic, this collaborative, back-and-forth refinement process ensured careful cross-validation of themes.
We acknowledge that a single-author finalization step limits our ability to quantify agreement.
Future work could include double-coding a subset of studies to quantify reliability and further strengthen the coding process.
% A complete replication package (search strings, study list, raw data extracts, taxonomy definitions, and diagrams) is publicly available, allowing others to audit and extend our work.

\subsection{Construct Validity}

Construct validity concerns whether the operational measures in this SLR truly capture the concepts we intend to study. 
Key threats include 
(a) selection bias in how we defined and applied inclusion/exclusion criteria, 
(b) instrumentation bias in data extraction, and 
(c) ambiguity in how key terms---such as runtime composition in SoS contexts---are interpreted across research communities. To mitigate these threats, we: 

\begin{itemize}
    \item Defined core concepts (e.g., runtime composition, interoperability, adaptation, dynamic reconfiguration) based on foundational literature, before data extraction.
    \item Developed a standardized codebook with clear definitions for each data item (D\textsubscript{1}--D\textsubscript{10}), ensuring that constructs such as ``challenge themes'' and ``tool categories'' were consistently interpreted across studies.
    \item Pilot-tested our data-extraction protocol on a random sample of 10 primary studies, refining ambiguous codes before complete extraction.
    \item Employed dual independent extraction: two reviewers extracted data in parallel, then reconciled discrepancies through discussion.
\end{itemize}

\subsection{Conclusion Validity}

Conclusion validity pertains to the reliability of inferences drawn from the data. 
Potential threats include 
(1) researcher bias in TA, and (2) over-generalization of findings. 
We mitigated these by:
\begin{itemize}
    \item Using a structured data extraction form to gather consistent information across all studies.
    \item Maintained detailed records of our decision-making processes throughout the analysis.
    \item Applied established qualitative analysis techniques (TA).
    \item Triangulated findings across multiple sources (e.g., cross-referencing identified challenges with proposed solutions).
    \item Provided extensive traceability between our findings and the primary studies through comprehensive referencing.
    % \item Made our complete dataset available in the replication package to enable independent verification.
\end{itemize}

\section{Related Work}
\label{sec:related-work}

This section examines prior systematic reviews, mapping studies, and surveys related to this study. 
Table~\ref{tbl:comparison_existing_reviews} provides a detailed comparison between our SLR and existing reviews across key dimensions, highlighting our distinct contributions.

Our SLR builds upon and extends existing literature in several important ways. While previous reviews have focused on specific aspects of SoSE, our work offers a unified taxonomy that integrates multiple dimensions, including runtime composition challenges, tools, evaluation methods, and domain applications.

\begin{table*}[t]
\centering
\footnotesize
\caption{Comparison of results between our SLR and existing secondary studies}
\label{tbl:comparison_existing_reviews}
\begin{tabular}{%
  >{\raggedright\arraybackslash}p{0.11\textwidth}  % Study
  >{\raggedright\arraybackslash}p{0.17\textwidth}  % Scope
  >{\raggedright\arraybackslash}p{0.27\textwidth}  % Key Findings
  >{\raggedright\arraybackslash}p{0.16\textwidth}  % Overlap
  >{\raggedright\arraybackslash}p{0.24\textwidth}  % Distinct Contributions
}
\rowcolor[rgb]{0.90,0.90,0.90}
\hline
\textbf{Study} & \textbf{Scope} & \textbf{Key Findings} & \textbf{Overlap} & \textbf{Distinct Contributions} \\
\hline
\citep{10.5555/3492252.3492254}
& Architectural patterns in SoS (Centralized, SOA, MSA).
& Pattern-based framework (HSH-SoS) categorizing major SoS design styles.
& Confirms prevalence of SOA/MSA patterns.
& We extend beyond patterns to runtime-composition challenges and tools. \\
\hline
\citep{9593274}
& Factors influencing architecture selection under dynamic evolution.
& Emphasis on interdependency, autonomy, and uncertainty.
& Agrees that dynamic evolution demands adaptive architectures.
& We map specific evaluation methods across themes. \\
\hline
\citep{8836749}
& Model-based frameworks (UAF + V-Model) for handling dynamic changes.
& Classification of modeling techniques for SoS development.
& Shares interest in modeling and V\&V.
& We offer a unified taxonomy integrating modeling with resilience, tools, and analytics. \\
\hline
\citep{10.1145/3519020} 
& Evaluation methods and metrics for SoS architectures.
& Cataloged evaluation methods, metrics, and trade-offs in SoS architecture assessment.
& Reinforces our finding on simulation dominance in evaluation.
& We embed evaluation within a broader SoSE framework (incl.\ tools, domains). \\
\hline
\citep{10.1145/3659098}  & Interoperability types in systems. & Identified diverse interoperability types and shift toward socio-technical concerns & Interoperability challenges in complex systems. & We incorporate interoperability within a broader taxonomy of SoS challenges and map specific solutions for runtime composition. \\

\\\hline
\citep{s24092921} & AI methods for dynamic interoperability. & Covered AI approaches like ontology alignment, heterogeneous data inference, and neural transcoding. & Underrepresentation of AI-driven approaches for interoperability. & We situate AI approaches within a comprehensive mapping of all solution types for runtime composition in SoS. \\
\hline
\citep{NAUDET2023695} & Cognitive interoperability in CPS enterprises. & Identified gaps in maturity models and the need for human-machine collaboration. & Cognitive aspects of system integration & We map cognitive interoperability within the broader context of runtime composition challenges in SoS. \\
\hline
\citep{10.1145/3487921} & Uncertainty in self-adaptive systems. & Explored community perceptions and approaches for managing unanticipated changes. & Explored uncertainty challenges & We integrated these challenges in the broader taxonomy. \\
\hline
\citep{10178527} & Digital twins in SoS. & Found certain digital twin challenges that need further study. & Confirms digital twins as an SoS integration approach. & We analyze digital twins as one of several solution approaches within a comprehensive framework for runtime composition. \\
\hline
\citep{9812681} & Design patterns in SoS & Synthesized a catalog of SoS design patterns. & Confirms SOA as a key design pattern for SoS. & Our SLR examines patterns alongside other solutions to enable runtime composition and adaptation. \\
\hline
\citep{9306923} & IoT in SoS contexts. & Analyzed methodologies, frameworks, and tools for IoT-SoS integration & Confirms IoT as one of the key domains for modern dynamic SoSs. & We  provide a domain-agnostic analysis of runtime composition challenges and solutions. \\
\bottomrule

\end{tabular}
\end{table*}

\subsection{Architecture and Modeling Approaches}

Previous architectural studies have explored various patterns and frameworks for SoS design. \cite{10.5555/3492252.3492254} proposed the HSH-SoS framework, categorizing major architectural styles (Centralized, SOA, MSA), while \cite{9593274} analyzed factors influencing architecture selection under dynamic evolution. In the modeling domain, \cite{8836749} classified techniques using UAF and V-Model frameworks for handling dynamic changes.

% While valuable, these architectural and modeling studies typically address isolated aspects of SoSE. Our taxonomy integrates these architectural concerns within a broader framework that maps specific methods and evaluation practices across themes, connecting architecture choices to runtime composition challenges and solution approaches.

While valuable, these architectural and modeling studies typically address isolated aspects of SoSE. Our taxonomy situates these architectural concerns within a broader classification that organizes runtime composition methods and their evaluation methods by thematic category, linking architectural choices to specific challenges and solutions.

\subsection{Evaluation and Interoperability}

Several reviews have focused on evaluation methods or interoperability challenges. \cite{10.1145/3519020} cataloged various methods and metrics for evaluating SoS architectures, reinforcing our finding on the dominance of simulation in such evaluations.
Interoperability studies by \cite{10.1145/3659098} identified diverse types and socio-technical concerns, while \cite{s24092921} and \cite{NAUDET2023695} explored AI methods and cognitive aspects of interoperability, respectively.

Our SLR situates these evaluation and interoperability concerns within a comprehensive mapping of solution approaches for runtime composition, providing context for when and how different methods apply across the SoS lifecycle.

\subsection{Emerging Approaches and Domain Applications}

\cite{10.1145/3487921} explored approaches for managing unanticipated changes in self-adaptive systems, \cite{10178527} analyzed digital twins as an integration approach, and \cite{9306923} investigated IoT-SoS integration. \cite{9812681} synthesized a catalog of SoS design patterns, confirming SOA's importance.

% Our work provides a domain-agnostic analysis that examines these emerging approaches as part of a unified framework for runtime composition and adaptation in SoS, mapping solutions to specific challenges across multiple dimensions.

% By integrating insights from these diverse studies, our SLR offers a comprehensive taxonomy that connects challenges, solutions, tools, and evaluation methods in SoSE, with particular emphasis on runtime composition---an aspect that cuts across prior review boundaries.

Our work provides a domain-agnostic analysis of emerging approaches to runtime composition and adaptation in SoS, integrating insights from diverse studies to offer a comprehensive taxonomy that maps challenges, solutions, tools, and evaluation methods across multiple dimensions, with particular emphasis on runtime composition---an aspect that cuts across prior review boundaries.

\section{Conclusion}
\label{sec:conclusion}

\updated{
% This SLR systematically analyzes the runtime composition landscape for dynamic SoSs by reviewing 80 studies published between 2019 and 2024.

This SLR examined the landscape of runtime composition in dynamic SoSs, analyzing 80 studies published between 2019 and 2024. Our goal was to synthesize current knowledge on challenges, solution approaches, tools, and evaluation methods to provide a coherent understanding of runtime-adaptive SoSs.

% We identified four primary challenge themes: (1) Modeling \& Analysis, (2) Robust \& Resilient Operations, (3) System Orchestration, and (4) Heterogeneity of CSs. 
% We also catalogued over 50 specific technical obstacles (e.g., emergent behavior prediction, workflow indeterminacy, black-box component integration). 
% We then organized proposed solutions into seven thematic areas: co-simulation \& digital twins, semantic ontologies, integration frameworks, architectural languages \& SOA, middleware \& communication, formal \& analytical methods, and resilience \& AI-driven approaches. 

We identified four primary challenge themes---Modeling \& Analysis, Robust \& Resilient Operations, System Orchestration, and Heterogeneity of CSs---comprising over 50 specific technical obstacles, including emergent behavior prediction, workflow indeterminacy, and black-box component integration. 
Proposed solutions clustered into seven thematic areas: co-simulation \& digital twins, semantic ontologies, integration frameworks, architectural languages \& SOA, middleware \& communication, formal \& analytical methods, and resilience \& AI-driven approaches.

% Key findings reveal that interoperability, heterogeneity, and the autonomy-coordination paradox are central challenges, while solutions emphasize SOA, semantic technologies, and adaptive frameworks. The EAF and simulation tools like Mosaik and Gazebo dominate the tooling ecosystem, yet fragmentation persists between design-time models and runtime dynamics.

% Our analysis underscores the need for holistic approaches that bridge technical and socio-technical dimensions, such as governance frameworks and AI-driven adaptation strategies. Despite advancements, gaps remain in benchmarking, scalability validation, and explicit treatment of directed/collaborative SoS types. 

% These findings provide researchers and practitioners a structured understanding of the current landscape and future directions for enabling runtime composable interoperable SoSs in increasingly complex and dynamic environments.

% Future research should prioritize standardized evaluation metrics, integrated toolchains, and interdisciplinary collaboration to advance runtime-adaptive SoS from theory to practice. 

Our analysis highlights that interoperability, heterogeneity, and the autonomy-coordination paradox remain central challenges, while SOA, semantic technologies, and adaptive frameworks dominate proposed solutions. Tooling is concentrated around platforms such as the EAF and simulation environments like Mosaik and Gazebo, yet fragmentation persists between design-time models and runtime dynamics.

These findings underscore the importance of holistic approaches that integrate technical and socio-technical dimensions, including governance frameworks and AI-driven adaptation strategies. Notably, gaps remain in benchmarking, scalability validation, and explicit support for directed and collaborative SoS types.

By providing a structured mapping of challenges, solutions, tools, and evaluation methods, this review offers researchers and practitioners a clear picture of the current state of runtime composable SoSs. Future research should focus on standardized evaluation metrics, integrated toolchains, and interdisciplinary collaboration to advance runtime-adaptive SoSs from theory to practice.

}

\section*{Acknowledgement}
\updated{We thank Dr. Rahul Mohanani and Dr. Nauman bin Ali for their helpful suggestions in improving the quality of this study.}

\appendix

\section{Acronyms}
\label{appendix:acronyms}
See Table~\ref{tbl:acronyms}.

\begin{table}[t]
\caption{List of acronyms}\label{tbl:acronyms}
\footnotesize
% \color{blue}
\begin{minipage}{\columnwidth}
\begin{tabular}{p{0.1\linewidth} p{0.75\linewidth}}
\toprule
Acronym & Full Form \\
\midrule
ADL & Architecture Description Language \\
API & Application Programming Language \\
BERT & Bidirectional Encoder Representations from Transformers\\
CPS & Cyber-Physical System \\
CS(s) & Constituent System(s) \\
CSoS & Combat SoS \\
CSSAD & CSoS autonomous decision-making \\
DAB & Discovery Access Broker \\
DCS & Discrete Controller Synthesis \\
DoSo & Degree of System Order \\
DSL & Domain-Specific Language \\
EAF & Eclipse Arrowhead Framework \\
IIoT & Industrial IoT \\
IoT & Internet of Things \\
ISoS & Infrastructure SoS \\
JSON & JavaScript Object Notation \\
MAPE-K & Monitor-Analyze-Plan-Execute over Knowledge \\
MPLS & Multiprotocol Label Switch \\
MQTT & Message Queuing Telemetry Transport \\
MSA & Micro-Service Architecture(s) \\
O/M independence & Operational and Managerial Independence \\
OPC UA & Open Platform Communications - Unified Architecture \\
OS & Operating System(s) \\
RDF & Resource Description Format \\
ReViTA & Reconfigurations Via Transient Architectural Configurations \\
RQ(s) & Research Question(s) \\
SiSoS & Software-intensive SoS \\
SLR & Systematic Literature Review \\
SOA & Service-Oriented Architecture(s) \\
SoS(s) & System(s) of Systems \\
SoSE & System of Systems Engineering \\
TA & Thematic Analysis \\
TSN & Time-Sensitive Networking \\
\bottomrule
\end{tabular}
\end{minipage}
\end{table}

\section{Selected Studies}
\label{appendix:selected-studies}
See Table~\ref{tbl:final-studies}.

\onecolumn
\footnotesize
\begin{longtable}
{m{0.025\textwidth}m{0.75\textwidth}m{0.15\textwidth}m{0.025\textwidth}}
    \caption{List of selected studies for the SLR.}
    \label{tbl:final-studies}\\
    \hline
    \rowcolor[rgb]{0.90,0.90,0.90}
    \textbf{ID} & \textbf{Study Title} & \textbf{Citation} & \textbf{Score} \\
    \hline
    \endfirsthead
    \multicolumn{4}{c}%
    {{\bfseries \tablename\ \thetable{}: continued from previous page}} \\
    \hline
    \rowcolor[rgb]{0.90,0.90,0.90}
    \textbf{ID} & \textbf{Study Title} & \textbf{Citation} & \textbf{Score} \\
    \hline
    \endhead
    \hline
    \multicolumn{4}{r}{{Continued on next page}} \\
    \endfoot
    \hline
    \endlastfoot

    \hypertarget{S01}{S01} & Applying model-based co-simulation on modular production units in complex automation systems. & \citep{binder2021applying} & 4 \\\hline
    
    \hypertarget{S02}{S02}   & Engineering interoperable, plug-and-play, distributed, robotic control systems for futureproof fusion power plants. & \citep{caliskanelli2021engineering} & 4 \\\hline

    \hypertarget{S03}{S03} & Machine learning-based discrete event dynamic surrogate model of communication systems for simulating the command, control, and communication system of systems. & \citep{kang2019machine} & 4 \\\hline

    \hypertarget{S04}{S04}  & Modeling challenges for Earth observing systems of systems & \citep{grogan2019modeling} & 5 \\\hline
    
    \hypertarget{S05}{S05}  
 & An interoperable tool-chain for energy monitoring applications. & \citep{brunelli2019interoperable} & 4 \\\hline
    
    \hypertarget{S06}{S06}  
 & MicroGraphQL: A unified communication approach for systems of systems using microservices and GraphQL. & \citep{de2022micrographql} & 5 \\\hline
    
    \hypertarget{S07}{S07}  
 & Towards an OPC UA Compliant Programming Approach with Formal Model of Computation for Dynamic Reconfigurable Automation Systems. & \citep{atmojo2019towards} & 3 \\\hline
    
    \hypertarget{S08}{S08}  
 & Autonomous and composable M\&S system of systems with the simulation, experimentation, analytics and testing (SEAT) framework. & \citep{mittal2020autonomous} & 4 \\\hline
    
    \hypertarget{S09}{S09}  
 & Building Systems of Systems with Escher. & \citep{canakci2021building} & 4 \\\hline
    
    \hypertarget{S10}{S10}  
 & Towards sustainable systems reconfiguration by an IoT-driven system of systems engineering lifecycle approach. & \citep{forte2022towards} & 3 \\\hline
    
    \hypertarget{S11}{S11}  
 & A resilience implementation framework of system-of-systems based on hypergraph model. & \citep{jiang2022resilience} & 3 \\\hline
    
    \hypertarget{S12}{S12}  
 & Native OPC UA handling and IEC 61499 PLC integration within the arrowhead framework. & \citep{cabral2020native} & 4 \\\hline
    
    \hypertarget{S13}{S13}  
 & Efficiently Composing Validated Systems Integration Gateways for Dynamic, Diverse Data. & \citep{rock2019efficiently} & 3 \\\hline
    
    \hypertarget{S14}{S14}  
 & Multi-scale hydrological system-of-systems realized through WHOS: the brokering framework. & \citep{boldrini2022multi} & 3 \\\hline
    
    \hypertarget{S15}{S15}  
 & A framework for the design of fault-tolerant systems-of-systems. & \citep{ferreira2024framework} & 3 \\\hline
    
    \hypertarget{S16}{S16}  
 & Adaptive Workflow of Service Oriented IoT Architectures for Small and Distributed Automation Systems. & \citep{zabasta2020adaptive} & 4 \\\hline
    
    \hypertarget{S17}{S17}  
 & Experiences of using linked data and ontologies for operational data sharing in systems-of-systems. & \citep{axelsson2019experiences} & 5 \\\hline
    
    \hypertarget{S18}{S18}  
 & Enabling architecture based co-simulation of complex Smart Grid applications. & \citep{binder2019enabling} & 5 \\\hline
    
    \hypertarget{S19}{S19}  
 & Evolving the ecosystem: Eclipse arrowhead integrates eclipse IoT. & \citep{kristan2022evolving} & 4 \\\hline
    
    \hypertarget{S20}{S20}  
 & Platooning LEGOs: An open physical exemplar for engineering self-adaptive cyber-physical systems-of-systems. & \citep{shin2021platooning} & 4 \\\hline
    
    \hypertarget{S21}{S21}  
 & From models to management and back: Towards a system-of-systems engineering toolchain. & \citep{kulcsar2020models} & 3.5 \\\hline
    
    \hypertarget{S22}{S22}  
 & Fuzzy architecture description for handling uncertainty in IoT systems-of-systems. & \citep{oquendo2020fuzzy} & 4 \\\hline
    
    \hypertarget{S23}{S23}  
 & IPSME-Idempotent Publish/Subscribe Messaging Environment. & \citep{nevelsteen2021ipsme} & 3 \\\hline
    
    \hypertarget{S24}{S24}  
 & System of systems lifecycle management---A new concept based on process engineering methodologies & \citep{kozma2021system} & 3 \\\hline
    
    \hypertarget{S25}{S25}  
 & Improving system-of-systems agility through dynamic reconfiguration. & \citep{fang2020improving} & 3.5 \\\hline

    \hypertarget{S26}{S26}  
 & Cautious adaptation of defiant components. & \citep{maia2019cautious} & 3 \\\hline  

    \hypertarget{S27}{S27}  
 & An ontological framework for opportunistic composition of IoT systems. & \citep{nundloll2020ontological} & 5 \\\hline 

    \hypertarget{S28}{S28}  
 & Research on Autonomous Reconstruction Method for Dependent Combat Networks. & \citep{sun2023research} & 3 \\\hline 

    \hypertarget{S29}{S29}  
 & A games-in-games approach to mosaic command and control design of dynamic network-of-networks for secure and resilient multi-domain operations. & \citep{chen2019games} & 4 \\\hline 

    \hypertarget{S30}{S30}  
 & Software mediators as first-class entities of systems-of-systems software architectures. & \citep{garces2019software} & 5 \\\hline 

    \hypertarget{S31}{S31}  
 & Conceptual, Mathematical, and Analytical Foundations for Mission Engineering and System of Systems Analysis. & \citep{raz2024conceptual} & 3.5 \\\hline 

    \hypertarget{S32}{S32}  
 & NLP-based generation of ontological system descriptions for composition of smart home devices. & \citep{zhang2023nlp} & 3 \\\hline 

    \hypertarget{S33}{S33}  
 & Continuous verification with acknowledged MAPE-K pattern and time logic-based slicing: A platooning system of systems case study. & \citep{song2023continuous} & 3.5 \\\hline 

    \hypertarget{S34}{S34}  
 & Integrating systems of systems with a federation of rule engines. & \citep{berezovskyi2024integrating} & 3.5 \\\hline 

    \hypertarget{S35}{S35}  
 & Incremental, distributed, and concurrent service coordination for reliable and deterministic systems-of-systems. & \citep{dhiah2020incremental} & 3 \\\hline 

    \hypertarget{S36}{S36}  
 & Providing a User Extensible Service-Enabled Multi-Fidelity Hybrid Cloud-Deployable SoS Test and Evaluation (T\&E) Infrastructure: Application of Modeling and Simulation (M\&S) as a Service (MSaaS). & \citep{mittal2023providing} & 3.5 \\\hline 

    \hypertarget{S37}{S37}  
 & Design process for system of systems reconfigurations. & \citep{petitdemange2021design} & 4 \\\hline 

    \hypertarget{S38}{S38}  
 & Adore: An adaptation-oriented requirement modeling approach for systems of systems. & \citep{maciel2019adore} & 5 \\\hline 

    \hypertarget{S39}{S39}  
 & Heterogeneous Network-Driven Data Fabrics to Enable Multi-Mission Autonomy.  & \citep{hawkins2022heterogeneous} & 3.5 \\\hline 

    \hypertarget{S40}{S40}  
 & Digital twins for collaboration and self-integration. & \citep{esterle2021digital} & 4 \\\hline 

    \hypertarget{S41}{S41}  
 & Continuous verification of system of systems with collaborative MAPE-K pattern and probability model slicing. & \citep{song2022continuous} & 3.5 \\\hline 

    \hypertarget{S42}{S42}  
 & Modeling the dynamic reconfiguration in smart crisis response systems. & \citep{belala2022modeling} & 4 \\\hline 

    \hypertarget{S43}{S43}  
 & HolonCraft--an architecture for dynamic construction of smart home workflows. & \citep{wang2022holoncraft} & 4.5 \\\hline 

    \hypertarget{S44}{S44}  
 & Simulation-Based Engineering of Heterogeneous Collaborative Systems---A Novel Conceptual Framework. & \citep{perivsic2023simulation} & 4 \\\hline 

    \hypertarget{S45}{S45}  
 & Extending the CST: The Distributed Cognitive Toolkit. & \citep{gibaut2020extending} & 3.5 \\\hline 

    \hypertarget{S46}{S46}  
 & Principled and automated system of systems composition using an ontological architecture. & \citep{elhabbash2024principled} & 4.5 \\\hline 

    \hypertarget{S47}{S47}  
 & Simulation integration platforms for cyber-physical systems. & \citep{neema2019simulation} & 5 \\\hline 

    \hypertarget{S48}{S48}  
 & A cloud-based development environment using HLA and kubernetes for the co-simulation of a corporate electric vehicle fleet. & \citep{rehman2019cloud} & 4 \\\hline 

    \hypertarget{S49}{S49}  
 & Interoperability analysis via agent-based simulation. & \citep{pescatore2024interoperability} & 3.5 \\\hline 

    \hypertarget{S50}{S50}  
 & A Maude-Based Formal Approach to Control and Analyze Time-Resource Aware Missioned Systems-of-Systems. & \citep{eddine2023maude} & 3.5 \\\hline 

    \hypertarget{S51}{S51}  
 & Self-Adaptation of Loosely Coupled Systems across a System of Small Uncrewed Aerial Systems. & \citep{chambers2024self} & 4 \\\hline 

    \hypertarget{S52}{S52}  
 & System of system composition based on decentralized service-oriented architecture. & \citep{derhamy2019system} & 5 \\\hline 

    \hypertarget{S53}{S53}  
 & Mission reliability modeling and evaluation for reconfigurable unmanned weapon system-of-systems based on effective operation loop. & \citep{chen2023mission} & 3 \\\hline 

    \hypertarget{S54}{S54}  
 & A middleware for automatic composition and mediation in IoT systems. & \citep{elhabbash2022middleware} & 4.5 \\\hline 

    \hypertarget{S55}{S55}  
 & Communication oriented modeling of evolving systems of systems. & \citep{harbo2021communication} & 3.5 \\\hline 

    \hypertarget{S56}{S56}  
 & The interoperability challenge: building a model-driven digital thread platform for CPS. & \citep{margaria2021interoperability} & 4 \\\hline 
    
    \hypertarget{S57}{S57}  
 & Modeling \& simulation of software architectures of systems-of-systems: An industrial report on the Brazilian space system. & \citep{neto2019modeling} & 5 \\\hline 

    \hypertarget{S58}{S58}  
 & Two-way integration of service-oriented systems-of-systems with the web of things. & \citep{zyrianoff2021two} & 4 \\\hline 

    \hypertarget{S59}{S59}  
 & Hardware/software self-adaptation in CPS: the CERBERO project approach. & \citep{palumbo2019hardware} & 4 \\\hline 

    \hypertarget{S60}{S60}  
 &
    JARVIS, A Hardware/Software Framework for Resilient Industry 4.0 Systems. & \citep{parri2019jarvis} & 3 \\\hline 

    \hypertarget{S61}{S61}  
 & Service-oriented reconfiguration in systems of systems assured by dynamic modular safety cases. & \citep{thomas2021service} & 3 \\\hline 

    \hypertarget{S62}{S62}  
 & On Combining Domain Modeling and Organizational Modeling for Developing Adaptive Cyber-Physical Systems. & \citep{sudeikat2022combining} & 3 \\\hline 

    \hypertarget{S63}{S63}  
 & Interoperability for industrial internet of things based on service-oriented architecture. & \citep{lam2021interoperability} & 4 \\\hline 

    \hypertarget{S64}{S64}  
 & System of System Strategy for Multi-Level Interoperability for Smart Cities. & \citep{loss2023system} & 5 \\\hline 

    \hypertarget{S65}{S65}  
 & CorteX: A software framework for interoperable, plug-and-play, distributed, robotic systems of systems. & \citep{caliskanelli2021cortex} & 4 \\\hline 

    \hypertarget{S66}{S66}  
 & Employing discrete controller synthesis for developing systems-of-systems controllers. & \citep{li2024employing} & 4 \\\hline 

    \hypertarget{S67}{S67}  
 & A Maude-Based Rewriting Approach to Model and Control System-of-Systems' Resources Allocation. & \citep{dridi2022maude} & 3.5 \\\hline 

    \hypertarget{S68}{S68}  
 & Systems of Digital Twins and Physical Systems: Interoperability, Decentralization, and Mobility in Robotic Applications. & \citep{girke2024systems} & 5 \\\hline 

    \hypertarget{S69}{S69}  
 & DynADL-Dynamic Architecture Description Language for System of Systems. & \citep{hristozov2024dynadl} & 5 \\\hline 

    \hypertarget{S70}{S70}  
 & A system of systems architecture for the internet of things exploiting autonomous components. & \citep{nikolopoulos2019system} & 4 \\\hline 

    \hypertarget{S71}{S71}  
 & Multi-swarm-based cooperative reconfiguration model for resilient unmanned weapon system-of-systems. & \citep{sun2022multi} & 3 \\\hline 

    \hypertarget{S72}{S72}  
 & Deep reinforcement learning-based resilience enhancement strategy of unmanned weapon system-of-systems under inevitable interferences. & \citep{sun2024deep} & 3 \\\hline 

    \hypertarget{S73}{S73}  
 & Generation of Adaptation Strategies for Dynamic Reconfiguration of a System of Systems. & \citep{lee2021generation} & 4 \\\hline 

    \hypertarget{S74}{S74}  
 & System reconfiguration ontology to support model-based systems engineering: Approach linking design and operations. & \citep{qasim2023system} & 4 \\\hline 

    \hypertarget{S75}{S75}  
 & A Brief Introduction into (De-)Coupling Lifecycles in Net-Centric Systems-of-Systems. & \citep{schutz2022brief} & 4 \\\hline 

    \hypertarget{S76}{S76}  
 & KIPO opportunities for interoperability decisions in systems-of-information systems in the domain of environmental management. & \citep{fernandes2017kipo} & 4 \\\hline 

    \hypertarget{S77}{S77}  
 &  Exploring system of systems resilience versus affordability trade-space using a bio-inspired metric. & \citep{chatterjee2021exploring} & 3 \\\hline 

    \hypertarget{S78}{S78}  
 & Using modeling and simulation and artificial intelligence to improve complex adaptive systems engineering. & \citep{tolk2022using} & 3 \\\hline 

    \hypertarget{S79}{S79}  
 & Architecture Description Method for Open Systems-of-Systems to Reduce Misunderstanding the Scopes of Managed Objects. & \citep{kobayashi2020architecture} & 4 \\\hline 

    \hypertarget{S80}{S80}  
 & A self-adaptive system of systems architecture to enable its ad-hoc scalability: Unmanned vehicle fleet-mission control center case study. & \citep{sadik2023self} & 4 \\\hline

\end{longtable}

\bibliographystyle{cas-model2-names}

\bibliography{main}

\end{document}